\documentclass[aps,prb,reprint,groupedaddress]{revtex4-1}
\usepackage[pdftex]{graphicx}
\usepackage{hyperref}
\usepackage{dcolumn,physics}
\usepackage{bm,comment}
\usepackage{color}

\begin{document}
\preprint{}
\title{Spin-orbit coupled Hubbard skyrmions}
\author{Ryo Makuta}
  \email{makuta-r@g.ecc.u-tokyo.ac.jp}
\author{Chisa Hotta}%
   \email{chisa@phys.c.u-tokyo.ac.jp}
     \affiliation{Department of Basic Science, University of Tokyo, 3-8-1 Komaba, Meguro, Tokyo 153-8902, Japan}
\date{\today}
\begin{abstract}
When the material phases exhibit topological quantum numbers,
they host defects protected by the nontrivial topology.
Magnetic skyrmions are such ``quantized" objects
and although many of them are metals
they had been most likely treated in theories as purely classical spin states.
Here, we show that the electrons described by the Hubbard model with strong spin-orbit couplings
can exhibit various nano- or flake-size skyrmions in their ground state.
Importantly, the conducting electrons forming Fermi pockets themselves
carry textured magnetic moments in both real and momentum space and on top of that,
possess Chern numbers in their energy bands.
This quantum and conducting skyrmion is related to small skyrmions observed
in atomic-layered compounds.
We clarify how the effective magnetic interactions and magnetic anisotropies are tied to
the spin-orbit coupling, and how they influence the stability of skyrmions beyond the phenomenology.
\end{abstract}
\maketitle
\section{Introduction}
Skyrmions are topologically protected classical defects first discovered in a nonlinear field theory for mesons
in high-energy physics\cite{Skyrme1961}.
These studies are followed afterward by the intent discussions in the 1990s,
regarding the charged excitations of quantum Hall ferromagnets as skyrmions\cite{Sondhi1993,Tycko1995}.
The latest analog appears in magnetic materials \cite{Bogdanov1989-2}
particularly, in a family of non-centrosymmetric B20-type transition-metal silicides and germanides
\cite{Muhlbauer2009,Neubauer2009,Yu2010,Yu2011,Seki2012,Tokunaga2015},
which have been experimentally established as platforms of skyrmions that extend from nanometers to micrometer scale.
In these chiral ferromagnets with broken inversion symmetry,
the classical descriptions based on field theories have successfully explained the phenomenological features
of their swirling magnetic structures nowadays called Bloch-type\cite{Nagaosa2013};
the underlying helical magnetic structures running in different directions are combined and
thermodynamically converted to skyrmions at a very particular range of finite temperature and magnetic field strength.
Later, another series called N\'eel-type skyrmions was found in the cycloidal magnets
in non-centrosymmetric polar crystals
\cite{Kezsmarki2015,Fujima2017,Butykai2017,Kurumaji2017}.
Meanwhile, some other classes of skyrmions in centrosymmetric materials\cite{Kurumaji2019,Hirschberger2019}
show that the frustrated classical magnetic interactions \cite{Okubo2012}
or RKKY interactions\cite{Hayami2017,Wang2020,Mitsumoto2021} can also bear very similar textures to Bloch-type ones.
\par
These ``classical skyrmions" are so far believed to be well understood by the classical-spin-based Ginzburg-Landau theories or models \cite{Nagaosa2013,Yi2009,Buhrandt2013},
which seemingly suggests it be buried in oblivion that most of the skyrmionic materials discovered so far are metals.
Importantly, for skyrmions found in laboratories,
the electrons itinerate as charge carriers which themselves carry spatially textured magnetic moments.
This means that although the ordered moments of long wavelength may fit the classical description,
their underlying origins should be far from being purely classical.
Previously, the magnetic interactions discussed in classical theories were
ferromagnetic Heisenberg and Dzyaloshinskii-Moriya (DM) interactions as well as magnetic anisotropies
which are treated as handweaving free parameters allowed from the crystal symmetry point of view;
it remains unsettled how they stem from both spin-orbit coupling (SOC) and inversion-symmetry breaking.
\par
The current challenge is to truly understand the SOC as the direct origin of skyrmions visibly formed by {\it conducting} electrons.
It is found that the SOC Hubbard model can afford metallic skyrmions that carry both charges and spin moments
while maintaining its topological feature as Chern numbers on top of their magnetic quantum winding numbers.
Quite importantly, the observed skyrmions are different from the Bloch or N\'eel type
sizable skyrmions previously reported at finite temperatures.
They are either nano- or flake-size and appear in the ground state
for a wide range of Coulomb interactions when the SOC is substantially large.
It turns out that the DM interactions and the types of spin anisotropies for these insulating magnets
are not at all independent parameters, being strictly governed by the SOC.
Our conclusions urge the reconsideration of the models and conclusions drawn from classical theories.

\vspace{5mm}
\leavevmode \\
\section{Model and method}
\begin{figure*}
  \includegraphics[width=1.0\textwidth]{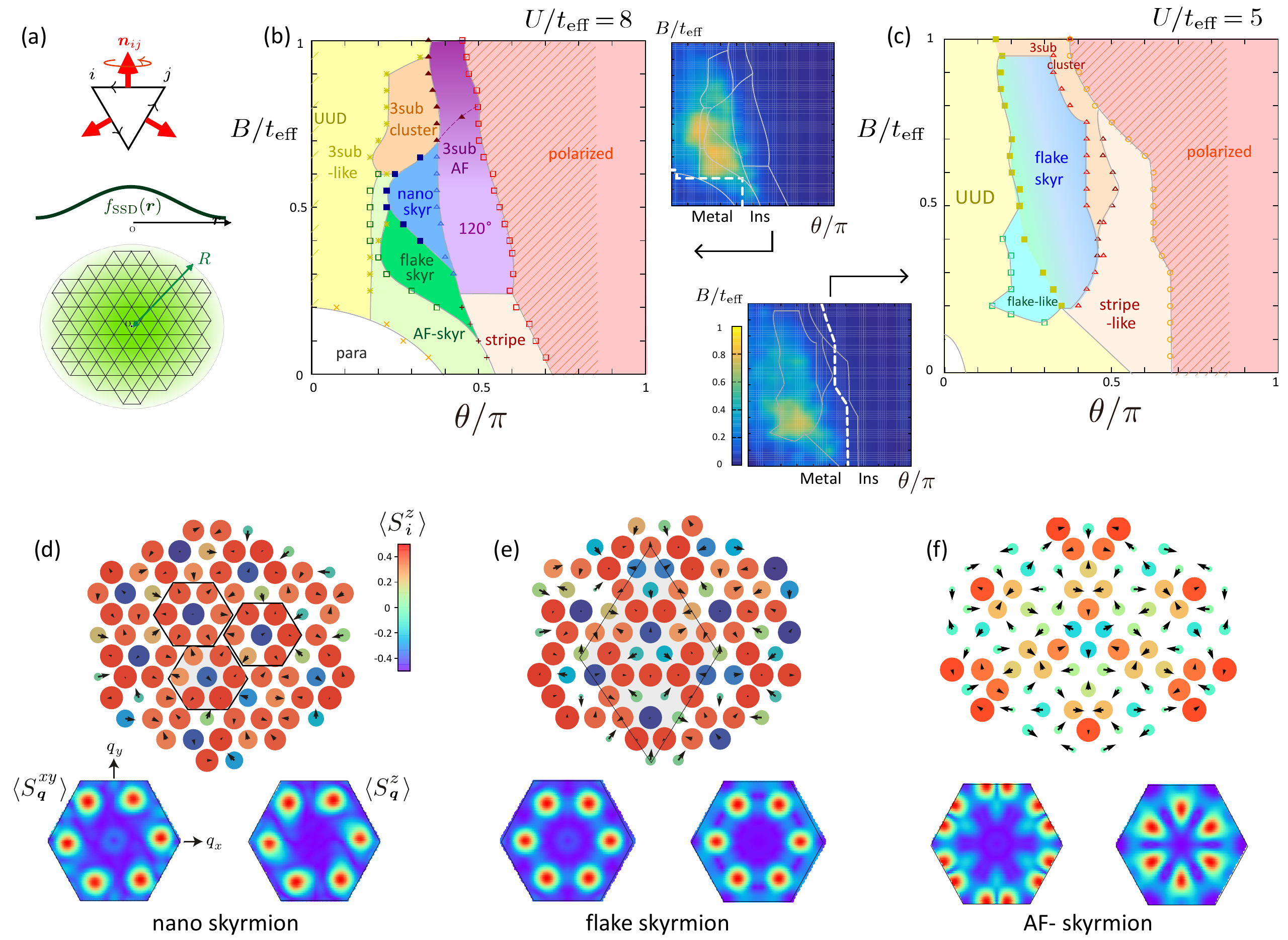}
  \caption{(a) Schematic illustration of the rotation axis $\bm n_{ij}$ included in the
   SU(2) gauge field $U_{ij}$ when electrons hop from site -$j$ to site-$i$.
  $N=90$ finite size cluster with SSD boundary conditions mainly used in the present calculation and
  the envelope function $f_{\rm SSD}(\bm r)$ (see Model and Method).
  (b,c) Ground state phase diagram of Eq.(\ref{eq:ham}) at $U/t_{\rm eff}=8$ and 5.
  The two small panels give the density plots
  of the quantum winding number $W$ and the metal-insulator phase boundaries for the two diagrams.
  (d,e,f) Realspace spin configuration of the SSDMF solutions for the nano-, flake- and tetrahedral AF-skyrmions.
  The arrows represent $xy$-component of each spin and the color and size of the circle represent $z$-component.
  The lower two panels are the spin structural factors $\langle \bm S_{\bm q}\rangle$
  in momentum space for the in-plane $xy$(left) and $z$-components(right panel).
}
  \label{f1}
\end{figure*}

We consider the Hubbard model at half-filling with Rashba SOC on the triangular lattice
whose Hamiltonian is given as
\begin{align}
\hat{\mathcal{H}}=&
-\sum_{\langle i,j\rangle}
\bm{c}_{i}^\dagger \big( t +i\lambda
(\bm{n}_{ij}\cdot\bm{\sigma})\big) \bm{c}_{j} + \mathrm{h.c.}
\nonumber\\
& +U \sum_{j}  \hat{n}_{j\uparrow}\hat{n}_{j\downarrow}
  -\sum_{j} \bm{c}_{j}^\dagger (\bm B \cdot \bm \sigma) \bm{c}_{j},
\label{eq:ham}
\end{align}
where $\langle i,j\rangle$ runs over all pairs of nearest neighboring sites,
$\bm c_{i}^\dagger= (c_{i \uparrow}^\dagger,c_{i \downarrow}^\dagger)$ is the creation operator of up and down spin electrons,
$n_{i}$ is an electron number operator,
and $U$ and $\bm B$ are on-site interaction and an external magnetic field, respectively.
The spin-dependent hopping integral $\lambda$ originates from the SOC\cite{Witczak2014,Nakai2022},
and when combined with the $t$-term we obtain the form,
\begin{equation}
t+i \lambda (\bm{n}_{ij}\cdot\bm{\sigma})= t_{\mathrm{eff}} \mathrm{e}^{i(\theta/2)\bm{n}_{ij}\cdot\bm{\sigma}}, \;\;
\end{equation}
using the SU(2) gauge field, $U_{ij}=\mathrm{e}^{i(\theta/2)\bm{n}_{ij}\cdot\bm{\sigma}}$,
where $\bm{\sigma}=(\sigma^{x},\sigma^{y},\sigma^{z})$ is the Pauli matrix,
$t_{\mathrm{eff}}=\sqrt{t^{2}+\lambda^{2}}$ and $\theta=2\mathrm{arctan}(\lambda/t)$.
The gauge field rotates the spin orientation by $\theta$ about $\bm{n}_{ij}$-axis
when the electron hops to the nearest neighbor sites as shown in Fig.~\ref{f1}(a).
\par
The direction of the rotation axis $\bm{n}_{ij}$ is determined by the crystal symmetry of the materials.
For example, in the B20-type chiral magnets with $P2_13$ or $P4_132$ space groups,
$\bm{n}_{ij}$ should be parallel to the bond directions and favor helical structures.
Whereas, for polar magnets of $C_{nv}$ group, $\bm{n}_{ij}$ should be perpendicular to bonds
and host cycloidal magnetic structures.
In the Hubbard model, the orientation of $\bm n_{ij}$ turns out to be an important factor
determining not only the kind of skyrmions but also its stability.
In our calculation, we consider $C_{3v}$ type $\bm{n}_{ij}$ placed inside the 2D plane (see Fig.~\ref{f1}(a)) and
$\bm B$ perpendicular to the plane
unless otherwise noted.
\par
The reason why Eq.(\ref{eq:ham}) remained unexplored
despite the importance of studying electronic models for skyrmions is purely because of technical difficulty.
Quantum Monte Carlo method (QMC) suffers a sign problem.
Other standard numerical solvers that can be applied to Eq.(\ref{eq:ham}) use the finite-size cluster
that often misfits with periods of incommensurate orders.
In such cases, the stability of the true ground state is underestimated or overlooked without knowing its periods
{\it a priori}.
For example, dynamical mean-field theory (DMFT) is one of the most successful solvers
for Hubbard models\cite{Metzner1989,Georges1992} and give a good description for Mott transition,
whereas it is built on a quality of self-energy included in the frequency-dependent Green's function
which does not take account of the longer-range fluctuations or spatially long-distance correlation effects.
When extended to larger-unit clusters known as c-DMFT\cite{Maier2005},
the results show substantial dependence on the size and shape of the clusters,
and are not suitable for the description of phases with possible magnetic structures or
correlations of large spatial periods.
In most cases, the benefit of using c-DMFT or other cluster-based methods overwhelms this disadvantage,
but for the present Hubbard model with SOC, the artifact hinders the unbiased search of skyrmions.
Recently, a theoretical treatment called sine-square-deformed mean-field theory (SSDMF)
has been developed\cite{Kawano2022} and was successfully applied to the same Hamiltonian as Eq.(\ref{eq:ham})
on a square lattice\cite{Kawano2023}, whose phase diagram turned out to host a variety of magnetic structures
including large-scale incommensurate spin-density-wave states, spiral, stripe, and vortex phases,
which seriously compete with each other.
There, the reliability of unbiasedness and quality of the SSDMF results
are confirmed through the comparison with four different analytical and numerical methods
some of which take full account of the correlation effects;
For example, c-DMFT predicted nonmagnetic insulating phases that contradict with the QMC studies,
whereas SSDMF agrees with QMC.
The two-fold periodic commensurate magnetic order reported in the c-DMFT phase diagram was also
an artifact; the density matrix embedding theory (DMET)\cite{Knizia2012},
which takes account of the full correlation and
can capture the large spatial period orders\cite{Plat2020,Kawano2020}
favor the SDW phase, in agreement with our SSDMF.
\par
Indeed, SSDMF is not simply one of the lowest approximation method\cite{Kawano2022};
The two utmost advantages beyond the simple mean field are,
it can get rid of the finite size effect,
and the ability to avoid the bias to wavevectors commensurate with the cluster size.
In Ref.[\onlinecite{Kawano2022}], it is even shown that the charge gap of a 1D and square lattice Hubbard model
evaluated by SSDMF agrees with those of the exact solutions.
This is due to the real-space renormalization effect caused by SSD\cite{Hotta2013}.
We thus apply the SSDMF to Eq.(\ref{eq:ham}) for the cluster shown in Fig.~\ref{f1}(a)
and successfully disclose the basic properties of the ground state.
The SSD combined with the quantum many-body solvers have established that it
can quantitatively accurately evaluate incommensurate orders or correlated density-wave states
in several quantum spin models and Hubbard models\cite{Hotta2012,Hotta2013}.
Although SSDMF relies on the Hartree-Fock approximation,
the predicting powers of SSDMF and SSD-DMRG about the most dominant charge and spin correlations
are almost comparable, and the results agree almost perfectly, which is shown in Appendix \ref{sec:dmrg}.
\par
\begin{figure}
  \includegraphics[width=0.45\textwidth]{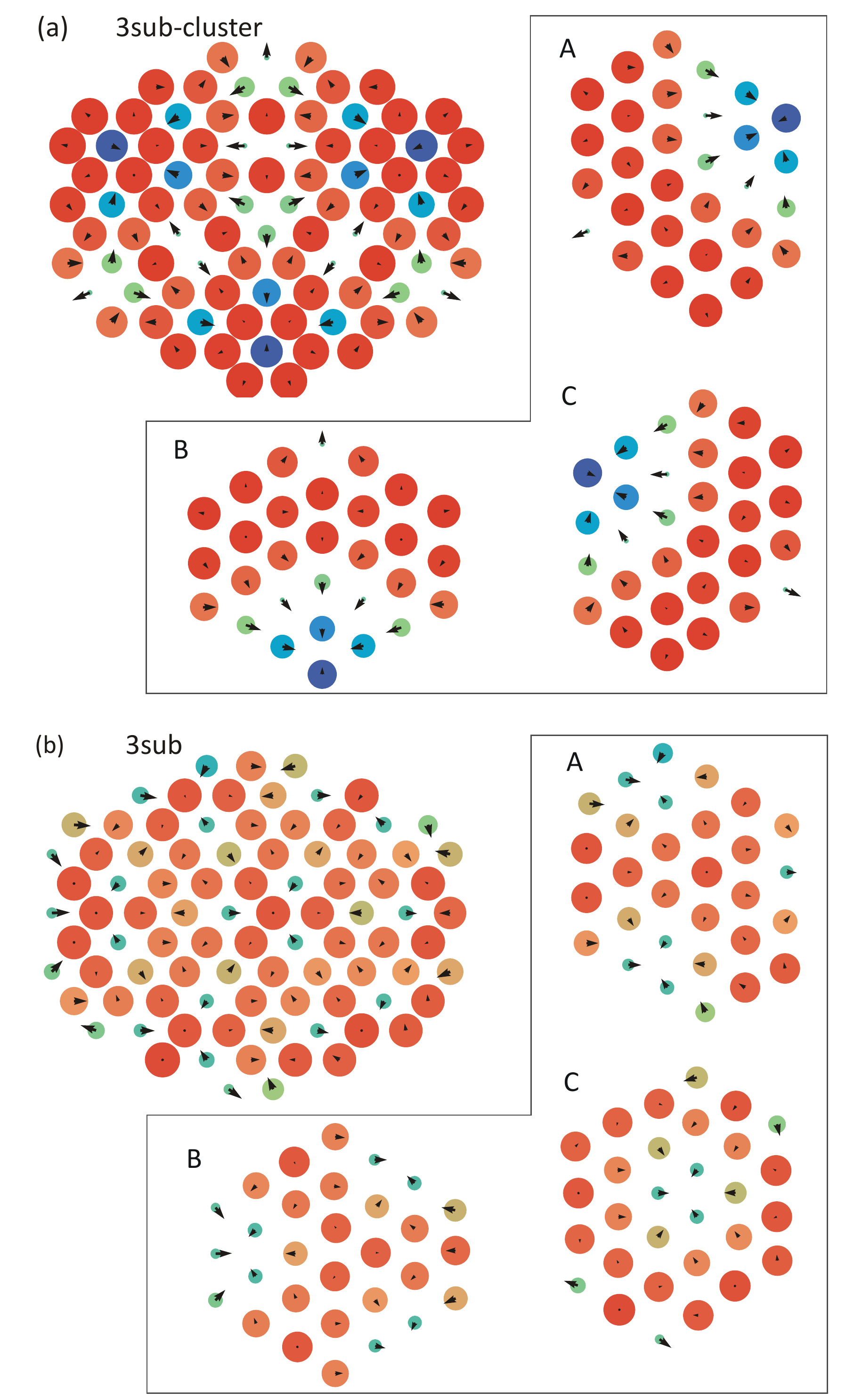}
  \caption{(a,b) Real space spin configuration of the SSDMF solutions for 3sub-cluster skyrmions found
   at $(U/t_{\rm eff},B,\theta/\pi)=(8,0.85,0.2),(5,1,0.3)$.
For both skyrmions, configurations on sublattices A, B, and C are plotted separately.
}
  \label{f3sub}
\end{figure}
\begin{figure}
  \includegraphics[width=0.45\textwidth]{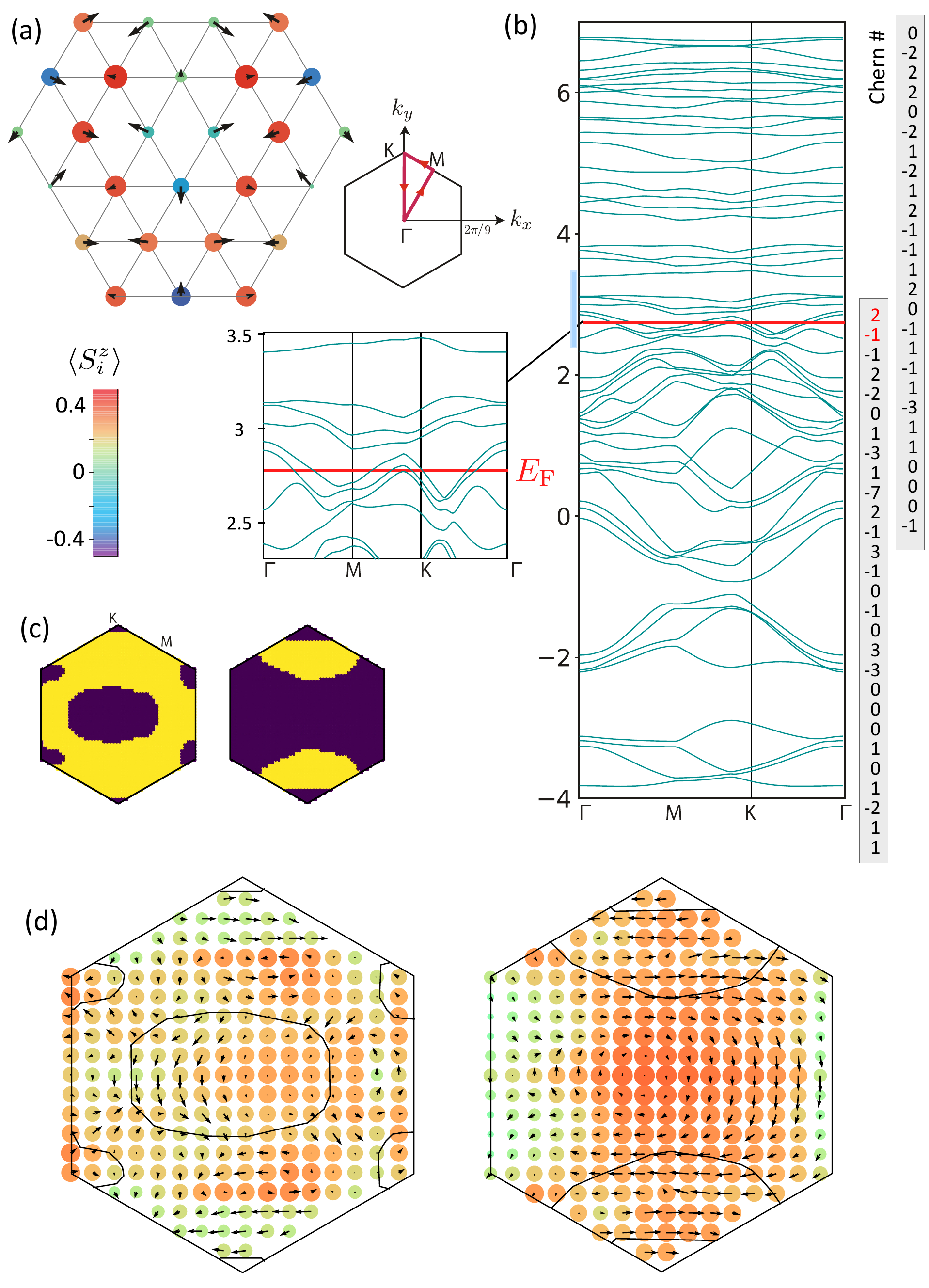}
  \caption{Metallic flake-skyrmion ground state obtained by SSDMF at $U/t_{\rm eff}=5$, $B/t_{\rm eff}=0.45$ and $\theta=0.35\pi$.
 (a) Spin configuration inside the 27-site unit.
 (b) Energy band structure consisting of 54 bands with Chern numbers is shown on the r.h.s.
  The inset shows the magnified bands near the Fermi surface.
 (c) Fermi surfaces of valence (hole) and conduction (electron) bands.
  The bright/dark regions are those below/above the Fermi level.
 (d) Spin textures in momentum space ($z$-component as density plots and $xy$-components as arrows)
    carried by the valence and conduction bands. The Fermi pockets are given as the guide to the eye. }
  \label{f2}
\end{figure}
We briefly outline the SSDMF theory\cite{Kawano2022}.
We prepare a finite-size hexagonal-like cluster and multiply an envelope function
to Eq.(\ref{eq:ham}) that has a sine-squared functional form gradually decreasing from the
maximum at the center ($\bm{r}_i=0$) toward the cluster edges (see Fig.~\ref{f1}(a)) given as
 $f_{\mathrm{SSD}}(\bm{r})=(1+\cos (\pi |\bm{r}|/R))/2$
with a radius $R={\rm max}_{i}|\bm{r}_i|+1/2$.
The cluster size adopted are $N=90$ in Fig.~\ref{f1}(a) and $N=198$
to confirm the convergence of the finite size effect.
\par
The site-dependent mean fields $\langle n_{i}\rangle$ and $\langle \bm S_{i}\rangle$ are introduced
and are determined self consistently.
Our mean-field Hamiltonian with SSD is given by $\mathcal{H}_{\rm MF} =\mathcal{H}_0+\mathcal{H}_{U,\rm MF}$ where
\begin{eqnarray}
  \mathcal{H}_{U,\rm MF} &=& U\sum_{i}f(\bm{r}_i)\qty(\frac{\ev{n_i}}{2}n_i-2\ev{\bm{S}_i}\cdot\bm{S}_i) + E_c,\\
  E_c &=& -U\sum_{i}f(\bm{r}_i)\qty(\frac{\ev{n_i}^2}{4}-\ev{\bm{S}_i}^2).
\end{eqnarray}
The obtained solutions are site-dependent, and when given the deformed Fourier transform as
$\langle{\bm S}_{\bm q}\rangle= \sum_i f_{\mathrm{SSD}}(\bm{r}_i) \langle \bm S_{i}\rangle
 e^{i\bm q \bm r_i}/ \sum_i f_{\mathrm{SSD}}(\bm{r}_i)$,
they yield an unbiased period that almost exactly reproduces that of the bulk limit.
This is in sharp contrast to the results obtained by the periodic boundary condition which show significant size dependence even when the cluster period is prepared as compatible with the periodicity of the expected orderings.
Previously,
for example,
 the period of $(q,q)$ ($q=(1-1/\sqrt{10})\pi)$
is found to be exactly detected using the $8\times 8$ cluster\cite{Kawano2022}.
We perform the inverse Fourier transformation on $\tilde{S}^{\mu}_{i}(\bm{k})$ and $\tilde{n}_i(\bm{k})$
with normal periodic boundary conditions of $N=192$ clusters to obtain spin and particle densities on a
uniform periodic lattice.
\vspace{5mm}
\\
\section{Results}
\subsection{Phase diagram}
We first show the phase diagrams on the plane of $\theta$ and $B$ for $U/t_{\rm eff}=8$ and 5 in Figs.~\ref{f1}(b) and ~\ref{f1}(c).
The $\theta=0$ case is the standard Hubbard model at half-filling;
when $U/t_{\rm eff}=8$, the system is in a paramagnetic phase at $B=0$,
which is transformed at $B/t_{\rm eff}\ge 0.2$ to the state with all the triangles having
two-up and one-down spin orientations, which is known as UUD phase in quantum magnetism.
This result is consistent with the corresponding previous theories on the half-filled Hubbard model
\footnote{In standard mean-field approximation the magnetic insulators
are over stabilized so that the UUD state shall appear
at $B=0$ in both diagrams. However, SSDMF safely captures the paramagnetic ground state observed in previous studies
taking account of the correlation effects at their maximum, indicating that it includes the higher-order correlation effects
from the idea of energy renormalization done by SSD\cite{Hotta2013}.
Indeed the 120$^\circ$-based states appear in our phase diagram at the comparable $U/t\sim 10$
as Ref.[\protect\onlinecite{Shirakawa2017}].}
showing the spin liquid Mott insulator at $U/t\sim 8-10$ (while it is beyond our description)
and 120$^\circ$ N\'eel ordering at $U/t\gtrsim 10$
\cite{Morita2002,Yoshioka2009,Shirakawa2017}.
At $\theta=\pi$ with $t=0$ and $\lambda=1$, an almost fully polarized ferromagnet appears,
whose origin will be discussed shortly.
In the shaded area of the ferromagnetic phase, the metastable solution reminiscent of the N\'eel-skyrmion appears
which is discussed in Appendix \ref{sec:appmagstates}.
\par
At intermediate $\theta$, $t$ and $\lambda$ compete and generate various phases.
The most prominent feature is the emergence of
two small-size skyrmion phases at around $0.2\lesssim \theta/\pi \lesssim 0.5$,
which we call nano- and flake-skyrmions
based on the similarity with the previously found skyrmions in quantum spin models\cite{Sotnikov2021,Lohani2019};
their real space spin distributions and spin structural factors in the reciprocal space are shown
in Figs.~\ref{f1}(d) and ~\ref{f1}(e).
They have site-centered and bond-centered characters with 7-site and 27-site per unit, respectively.
The spin structural factors have peaks positioned at $\bm q=(0.955\pi,0)$ and $(0.955\pi,0.064\pi)$, respectively.
In the lower field region, the AF-skyrmion based on three sublattice structures of a 27-site unit is observed;
the structural factors show peaks at different points from the nano- and flake-skyrmions.
All of them are characterized as triple-$q$ states.
\par
For classical skyrmions, a winding number is used to characterize the slowly varying continuous spin structures,
which takes either an integer or half-integer numbers when the gradients of spins are integrated over the skyrmion unit.
In quantum systems, the spin moments fluctuate and squeeze on discrete lattice sites
and do not show global continuous structures.
The replacement of the winding number is given by the skyrmion number $W$ introduced in a Ref.[\onlinecite{Lohani2019}]
as the total solid angle of triangular spins as
\begin{eqnarray}
  W &=& \frac{1}{4\pi}\sum_{\langle ijk \rangle}^{\Delta}
  \atan2\bigg(  \frac{\langle \bm{S}_{i}\cdot \bm{S}_j\times \bm{S}_k\rangle}{S^3},
   \nonumber \\
   && \rule{10mm}{0mm} 1+\frac{\langle \bm{S}_i\cdot \bm{S}_j\rangle
   +\langle \bm{S}_j\cdot \bm{S}_k\rangle+\langle \bm{S}_k\cdot \bm{S}_i\rangle}{S^2}
   \bigg),
\end{eqnarray}
where the summation is taken over the skyrmion unit for triangles $\langle ijk \rangle$ with indices aligned counterclockwise.
The value of $W$ is not quantized and is relatively small\cite{Maeland2022}.
However, it can overall capture the nano- and flake-skyr phases as shown in the insets of Fig.~\ref{f1}.
\par
Another notable feature is the 3sub-cluster phase realized at large $B$ and $\theta\sim 0.2-0.3\pi$ whose magnetic structures are shown
in Fig.~\ref{f3sub}(a) and \ref{f3sub}(b), corresponding to $U/t_{\rm eff}=8$ and $5$, respectively.
At first sight, their underlying three-sublattice structures are not visible.
However, if we separately plot them for each sublattice in the right panels,
we find that each sublattice hosts a skyrmion with the 90-sites unit.
Similar three-sublattice skyrmion was experimentally found in MnSc$_2$S$_4$\cite{Gao2017,Rosales2020},
which are analyzed theoretically for the $J_1$-$J_2$-$J_3$ model\cite{Rosales2022} and DM model\cite{Rosales2015}.
Interestingly, this skyrmion is a source of the thermal Hall effect based on SU(3) magnons
characteristic of a three sublattice structure\cite{Takeda2024}.
Our results point out the possibility of realizing a similar structure
at the electronic level for the first time in the non-centrosymmetric Hubbard model.
\par
For other phases at $\theta/\pi \sim 0.5$ with increasing the magnetic field,
the system hosts stripes and  120$^\circ$- or umbrella-like three-sublattice phases both denoted as
``3sub AF". The details of these magnetic structures are given in Appendix \ref{sec:appmagstates}.
\par
Because SSDMF is based on the Hartree-Fock approximation, which generally over stabilizes
magnetic orderings or metal-insulator transitions, one might suspect that
a variety of phases that appear in the phase diagram may be wiped out when a strong fluctuation
due to higher order correlation is seriously included\cite{LeBlanc2015}.
However, in Ref.[\onlinecite{Kawano2023}], for the same SOC Hubbard model with small $U$,
it was shown that the SDW phase in the SSDMF phase diagram agrees with the DMET result
near the phase boundary, $U\sim 2$, and further agrees with the prediction of random phase approximation (RPA).
\par
To support the present phase diagram, we performed an RPA study for $ \theta = 0.3\pi$ and compared
it with our SSDMF solution at $U=5$, where the magnetic structure may be relatively fragile.
As we show in Appendix~\ref{sec:rpa}, the RPA and SSDMF show good agreement about the
magnetic structure factor.
Because $U=5$ is already sufficiently larger than $U_c$ of RPA, the magnetic orderings found in SSDMF
can be safely expected.
\par
It is an established knowledge that the standard mean-field calculation overestimates magnetic orderings.
Alongside, SSDMF is confirmed to capture accurately the wave numbers of orders without bias.
Combining these two facts, although the transition points or the amplitudes of magnetic orderings
shall not be quantitative, the SSDMF will not miss the magnetic orderings if really present
in the fully correlated case although not being possible to elucidate them by other methods.
The states not captured by SSDMF are the quantum nonmagnetic ones like spin liquid or valence bond crystals
which are not expressed by the standard Hartree-Fock approach to electrons.
Such states may sometimes appear in frustrated systems like the present triangular lattice
due to strong quantum fluctuation, while they are clearly out of the scope of the present study.
\subsection{Metallic skyrmions}
The phase diagrams have insulating and metallic regions whose boundaries are indicated by the broken lines
in the small panels of Fig.~\ref{f1}(b,c).
In our framework, the flake-skyrmion at $U/t_{\rm eff}=5$ can naturally appear as metals.
\par
In the metallic phases, the magnetic moments shrink,
and it becomes much more difficult to have a rigid regular magnetic structure compared to the insulating case.
In our calculation particularly at $U/t_{\rm eff}=5$,
the UUD, skyr-like, and stripe-like phases surrounding the flake-skyrmion phase
often show various numerical solutions with domains or irregular winding structures of spins.
Contrastingly, the flake-skyrmion metal is rigid (see Fig.~\ref{f2}(a)).
\par
Let us briefly show how we classify the states to be metallic or insulating.
Once we obtain the spatial periods of the spin structures by the dominant peak positions $S_\gamma (\bm q)$,
we can construct a system with periodic boundary conditions of the size of that period.
Based on the corresponding set of mean fields that work as one-body potentials,
the energy band structure and the Fermi surface are obtained,
which are shown in Figs.~\ref{f2}(b) and \ref{f2}(c) for flake-skyrmion.
The two bands cross the Fermi level which gives two hole pockets and electron pockets in different regions of the Brillouin zone.
Since not only the valence and conduction bands as well as other bands do not contact with each other,
we can calculate the Chern numbers as
\begin{equation}
C_n= \frac{1}{2\pi}\sum_k {\mathcal Im} \ln \Delta U_l(k), \;\;
\Delta U_l(k)=\prod_{plaquette} \langle u_k |\nabla_k | u_k \rangle ,
\end{equation}
for the $n$ th band using the fidelity of Bloch wave function $\Delta U_k$ along the discretized plaquette of
the reciprocal space\cite{Fukui2005}.
The results are shown on the r.h.s. of all the energy bands.
The conducting and valence bands have $C_n=2$ and $-1$, respectively.
Since the present system is gapless, the Chern number does not directly lead to the quantization of transport coefficients, but the underlying Berry curvature can bear an anomalous Hall conductivity. This effect can be described by the U(1) gauge field that is generated from the spin texture of the skyrmion.
For comparison, we also show the energy bands of the insulating AF-skyrmion and 3sub-cluster phases
with their Chern numbers in Appendix \ref{sec:appmagstates}.
\par
Due to SOC, the spins are no longer good quantum numbers.
Therefore, the eigenstates carry spin moments pointing in various directions;
we show in Fig.~\ref{f2}(d) the spin textures on valence and conduction bands of the flake-skyrmion.
Since the spin moments point almost opposite on different sides of the pockets,
one may expect the spin current to be observed if this state is realized.

\begin{figure}
  \includegraphics[width=0.45\textwidth]{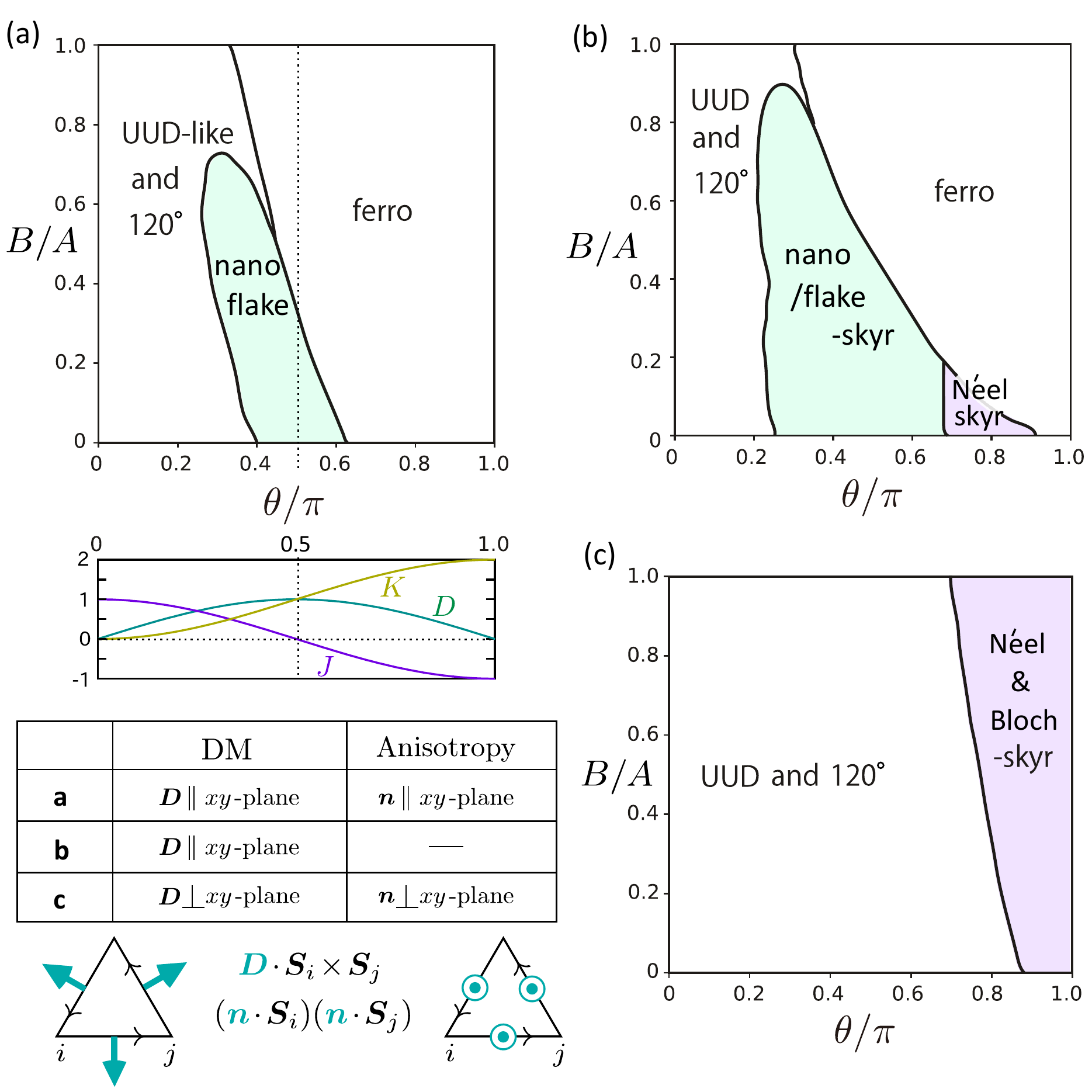}
  \caption{Phase diagram of the classical spin model including the Heiseberg interaction
$J=A\cos\theta$, DM interaction $D=A\sin\theta$ and spin anisotropy term $K=\sqrt{J^2+D^2}-J$,
taking $t_{\rm eff}=1$ and $U=10$ i.e., $A=4t_{\rm eff}^{2}/U=0.4$.
Panel (a) corresponds to the strong coupling limit of our SOC Hubbard model given in Eq.(\ref{eq:h2})
with in-plane $\bm D$ and $\bm n$, (b) is the one obtained by taking $K=0$,
and (c) shows the case where $\bm n_{ij}$ in Eq.(\ref{eq:h2}) is pointing perpendicular to the $xy$-plane which gives
$\bm D$ and $\bm n \perp xy$-plane.
In the lower panel of (a), the variation of $J/A=\cos\theta$, $D/A=\sin\theta$ , $K/A=(1-\cos\theta)$, and the detailed descriptions of $\bm D$ and $\bm n$ for the three cases are summarized.
The direction of vector $n_{ij}$ for the DM and anisotropy terms are bond direction ($i\rightarrow j$) dependent.
}
  \label{f3}
\end{figure}
%
%
\subsection{Spin models at strong coupling limit}
In our calculation, the large N\'eel-type skyrmion previously found in experiments and
classical spin models of the same symmetry class is not observed
even for parameters far outside the presented phase diagrams.
To understand why some skyrmions appear for certain parameters and others do not,
establishing the connection of Hubbard models with classical spin models is useful,
since the skyrmions in the latter types of models are intensively studied.
\par
The effective Hamiltonian of the strong coupling region $U/t_{\rm eff}\gg 1$ of Eq.(\ref{eq:ham}) at $B=0$ can be obtained
by the perturbation about $t_{\rm eff}/U$.
The second-order Hamiltonian includes the Heisenberg interaction, DM interaction,
and the Kaplan-Shekhtman-Aharony-Entin-Wohlman (KSAEW) term\cite{Kaplan1983,Shekhtman1992,Shekhtman1993} as
\begin{equation}
  \mathcal{H}^{(2)}=\sum_{\langle i,j\rangle}
  (J\bm{S}_i\cdot\bm{S}_j + \bm{D}_{ij}\cdot\bm{S}_i\times\bm{S}_j+K(\bm{n}_{ij} \cdot\bm{S}_i)(\bm{n}_{ij}\cdot\bm{S}_j))
\label{eq:h2}
\end{equation}
where $J=A\cos\theta$, $\bm{D}_{ij}=\bm{n}_{ij} A\sin\theta$, $A=4t_{\rm eff}^2/U$, and $K=\sqrt{J^2+D^2}-J$.
Importantly, the DM vector $\bm D_{ij}$ and spin anisotropy vector both point toward the direction of the rotation axis of the
SU(2) gauge, $\bm n_{ij}$, in Eq.(\ref{eq:ham}).
\par
Here, $\bm S_j$ in $\mathcal{H}^{(2)}$ is a quantum spin-1/2 operator.
Whereas taking its large-$S$ limit and regarding Eq.(\ref{eq:h2}) as classical Hamiltonian of vector spins
still works to understand the magnetic nature of the original Hamiltonian;
it is shown in the square lattice SOC Hubbard model that
the magnetic structure at finite $U$ and agrees well with that of the classical ones\cite{Kawano2023}.
To obtain the classical ground state phase diagram,
we compare the classical energies Eq.(\ref{eq:h2}) of representative magnetic orders that were stabilized
in the SSDMF calculation as a function of $\theta$ and $B$.
Figure~\ref{f3}(a) shows the resultant $(\theta, B)$ ground state phase diagram,
where we set the units of the interactions as constant, $A=4t_{\rm eff}^2/U=0.4$
corresponding to $t_{\rm eff}=1$ and $U=10$.
The nano/flake-skyr phase
appears at around $\theta \sim 0.5\pi$ and $B\lesssim 0.8$,
and the regions $\theta\lesssim 0.4\pi$ and $0.6\pi \lesssim \theta$ are dominated by
the three-sublattice (UUD or 120$^\circ$) phase and polarized ferromagnetic phase, respectively.
Here, we prepared all representative states that appear in the Hubbard phase diagram
as candidates and compared their energies; 120$^\circ$, UUD, nano-, flake-(19 site unit), and AF-skyrmions
and large-skyrmion (Ne\'el as well as other types) with an energetically optimized periodicity.
For the nano/flake phases, the two compete while nano-skyrmion is relatively stable.
The large Ne\'el-skyrmion state was energetically subtle and did not appear as the lowest energy state.
This diagram is fully consistent with the SSDMF results for Eq.(\ref{eq:ham}).
\par
The variation of $J$, $|\bm D_{ij}|$, $K$ as functions of $\theta$
shown in the lower panel of Fig.~\ref{f3}(a) helps to understand the energetics.
At $\theta<0.5\pi$, the Heisenberg exchange is antiferromagnetic ($J>0$) which becomes ferromagnetic ($J<0$)
at $\theta>0.5\pi$, having maximum amplitudes at $\theta=0$ and $\pi$.
The DM interaction takes the maximum at $\theta=0.5\pi$,
while the $K$ term is always positive and monotonously increases; this term dislikes the
adjacent spins to point in a similar direction, particularly in parallel to $\bm n_{ij}$.
Since Fig.~\ref{f3}(a) has $\bm n_{ij}$ pointing in-plane,
the large DM interaction at $\theta\sim 0.5\pi$
induces the typical cycloidal variation of moments along the bond directions,
while comparably large $K$ favors the spins to vary quite rapidly in space.
Both cooperatively stabilize the nano/flake-skyrmion phase when $B$ is not too large.
\par
The large N\'eel skyrmion, which may appear if present at around $\theta/\pi=0.6-0.9$
is not favored here, because it suffers the substantial loss of KSAEW energy
compared to the fully polarized state.
The reason why the previous classical model could easily host large skyrmions
was that the anisotropy interactions (different from KSAEW) and $\bm D_{ij}$ are dealt with
independently as free parameters, which is not allowed in Hubbard models.
To confirm this scenario, we calculated the phase diagram by artificially setting $K=0$
as shown in Fig.~\ref{f3}(b). At a small magnetic field next to the ferromagnetic phase
the large N\'eel skyrmion is recovered.
For further reference, we show in Appendix~\ref{sec:clspin} the above-mentioned
cases like $\bm D\parallel xy$ and $\bm n_{ij}\perp xy$ which host N\'eel skyrmion,
while they are unphysical in considering the SOC Hubbard model as their origin.
\par
Nonetheless, since the large N\'eel skyrmion is relevant not only to classical models but also to materials,
it is natural to expect another channel to realize it.
We thus restart from the model Eq.(\ref{eq:ham}) with
$\bm n_{ij}$ pointing perpendicular to the two-dimensional $xy$-plane for all bonds.
This modification does not change the form of Eq.(\ref{eq:h2}),
and its classical phase diagram is shown in Fig.~\ref{f3}(c);
the nano/flake-skyrmions no longer exist while the large-skyrmion appears at large $\theta$ region
instead of the polarized ferromagnets.
The large skyrmion here equivalently favors N\'eel, Bloch, or their mixtures because
$\bm n_{ij}\cdot \bm S_i\times \bm S_j$ depends solely on the relative angles
of classical spins projected onto the $ xy$ plane, which is identical for the two skyrmions.
This situation is realized in Mn$_{1.4}$Pt$_{0.9}$Pd$_{0.1}$Sn where the mixture of N\'eel and Bloch structures is observed
\cite{Nayak2017,Peng2020}.
\par
At the fourth order of perturbation, the ring exchange interaction including SOC appears.
We have examined the amplitude of these terms in Appendix~\ref{sec:4th} and found that
they remain one order smaller than the other magnetic interactions.
While these higher-order fluctuations will influence the amplitude of spins,
their effect is small since the phase diagrams we saw in Figs.~\ref{f1}(b) and~\ref{f1}(c)
are similar, although the degree of fluctuation is by orders of magnitude larger for the latter.
\begin{table}[tbp]
\caption{
Details of non-centrosymmetric materials hosting skyrmions based on helical (he) and cycloidal (cy) magnets;
space group, metallic/insulating (M/I), and the size of the observed skyrmions.
Those belonging to B20-group ($P2_{1}3$ or $P4_{1}32$) are helical/conical magnets
that form Bloch skyrmions, and the corresponding DM interactions are $\bm D_{ij} \parallel$ bonds.
Here, EuPtSi is classified in a different column since it is an RKKY system
where $S=7/2$ localized magnetic moments are coupled to conduction electrons.
The $C_{nv}$, $R_{3m}$ ,and $P4_{cc}$ systems are cycloidal magnets that form N\'eel skyrmions,
and the corresponding DM interactions are $\bm D_{ij} \perp$ bonds which are pointing mainly in-plane.
The latter groups correspond to our model.
}
  \centering
  \begin{tabular}{llccrr}
    \hline
 &material & symm. & M/I & [nm] & ref \\
\hline
& MnSi  &  $P2_{1}3$  & M & 18 & [\onlinecite{Muhlbauer2009}] \\
& FeGe  &  $P2_{1}3$  & M & 70  & [\onlinecite{Yu2011}] \\
& MnSi$_{1-x}$Ge$_{x}$ & $P2_{1}3$ & M & 60 & [\onlinecite{Shibata2013},\onlinecite{Grigoriev2013}] \\
& MnGe & $P2_{1}3$& M & 3& [\onlinecite{Kanazawa2011}] \\
& Mn$_{1-x}$Fe$_{x}$Ge& $P2_{1}3$ & M &3-6& [\onlinecite{Tanigaki2015}] \\
(he)&    Cu$_2$OSe$_3$ & $P2_{1}3$ & I &62&[\onlinecite{Seki2012},\onlinecite{Adams2012}]\\
&    Fe$_{1-x}$Co$_{x}$Si & $P2_{1}3$ &  I & 30-230& [\onlinecite{Yu2010}]\\
&    Co$_{10-x/2}$Zn$_{10-x/2}$Mn$_{x}\!\!$& $P2_{1}3$ & M& 9 & [\onlinecite{Fujishiro2019}] \\
&    Co$_{8-x/2}$Fe$_x$Zn$_x$Mn$_4$& $P4_{1}32$ & M & 120 &[\onlinecite{Tokunaga2015},\onlinecite{Karube2018}]\\
&    FeCo$_{0.5}$Rh$_{0.5}$Mo$_3$N & $P4_{1}32$ & M & 110 & [\onlinecite{Li2016}]\\
    \hline
(he)&    EuPtSi   &$P2_{1}3$ & M &  1.8 & [\onlinecite{Kaneko2019},\onlinecite{Kakihana2018}]\\
    \hline
    \hline
(he/cy)&    Mn$_{1.4}$Pt$_{0.9}$Pd$_{0.1}$Sn &$D_{2d}$ & M & 150 &[\onlinecite{Nayak2017},\onlinecite{Peng2020}]\\
    \hline
    \hline
&    Fe/Ir &    $C_{3v}$ & M & 1 & [\onlinecite{Heinze2011}]\\
&    PdFe/Ir &  $C_{3v}$ & M &3 & [\onlinecite{Romming2013}]\\
&    Ir/Co/Pt & $C_{3v}$ & M & 30-90 & [\onlinecite{Moreau-Luchaire2016}]\\
(cy)&    CBST/BST & $C_{3v}$ & M &-- \;\;\;& [\onlinecite{Yasuda2016}]\\
&    SrIrO$_3$/BaTiO$_3$ & $C_{4v}$ & M&-- \;\;\;& [\onlinecite{Matsuno2016}]\\
&    SrRuO$_3$/BaTiO$_3$ & $C_{4v}$ & M&-- \;\;\;& [\onlinecite{Wang2018}]\\
    \hline
    \hline
&    GaV$_4$S$_8$ & $R_{3m}$  & I &17 & [\onlinecite{Kezsmarki2015}]\\
(cy)&    GaV$_4$Se$_8$& $R_{3m}$  & I &19 & [\onlinecite{Fujima2017},\onlinecite{Bordacs2017}]\\
&    VOSe$_2$O$_5$& $P4_{cc}$ & I &140 & [\onlinecite{Kurumaji2017}]\\
    \hline
  \end{tabular}
\label{tab1}
\end{table}
\section{Summary and Discussions}
We have so far clarified that the metallic and nano- or flake-size skyrmions can be realized in an
SOC Hubbard model at half-filling in a certain range of magnetic field perpendicular to the triangular lattice plane.
Key differences from the former theories are listed as follows;
Firstly, the conducting electrons forming Fermi pockets contribute to skyrmions.
They may remind of a spin-density wave state but are different in that
the spin momentum is not conserved on each band, namely,
each $k$-point at each energy band carries magnetic moment pointing in different directions.
Second, each band carries nonzero-integer Chern numbers originating from the U(1) gauge field created by the skyrmion structure which contributes to the Hall conductivity.
Finally, the major effective magnetic interactions well-defined in the Mott insulating state,
$J$, $D$ ,and $K$, are interrelated and are fully controlled by the
single parameter $\theta=2\arctan(\lambda/t)$, originating from the SOC.
Accordingly, the large skyrmions that were easily observed in idealized and phenomenological
classical spin models are not found.
\par
To make the discussions relevant to laboratory experiments,
we present in Table \ref{tab1} the representative non-centrosymmetric materials hosting skyrmions.
They are classified into separate columns by double-lines as
chiral magnets ($P2_13$, $P4_132$), atomic layers (Fe/Ir) or heterostructures (SrIrO$_3$/BaTiO$_3$ interfaces),
and bulk layered materials.
\par
The ones based on helical magnets (he) are known as B20-type structures with distorted cubic lattices.
The Bloch-skyrmions found there are mostly as large as 10-200 nm,
and are treated phenomenologically by the Ginzburg-Landau-based or lattice-based classical spin models;
the major interactions are the ferromagnetic Heisenberg and
DM interactions with $\bm D_{ij}$ pointing parallel to the bonds along the $\hat x,\hat y,\hat z$-directions
that favor helical or conical magnets.
These phenomenologies successfully reproduced the experimental phase diagram,
at finite temperatures within a certain field strength\cite{Nagaosa2013,Yi2009,Buhrandt2013}.
\par
However, in reality, most of these magnets are metallic;
for example, MnSi is a typical itinerant-electron magnet with
a concentration of moments being 0.4$\mu_B$ per Mn sites, which cannot be simply treated as a localized large-$S$ magnet.
Further, the microscopic origin of ferromagnetic interactions in such a situation is not clarified.
Whereas, EuPtSi having the same space group is different;
the Eu ions have well-localized $S=7/2$ that interacts with conducting $s$-electrons\cite{Kakihana2018,Kaneko2019},
which can be regarded as the RKKY systems.
The theories built on classical Kondo-lattice types of models naturally reconcile with this system
and exhibit phase diagrams with Bloch skyrmions\cite{Ozawa2017,Mitsumoto2021}
which are more or less the same as the ones commonly observed in B20-magnets.
Recently, the DFT calculations have shown that the mixed valence state of MnSi
can be separated into a relatively localized spin part and the conducting part\cite{Choi2019},
where they propose the Kondo-like model and show that the Kondo-coupling strength determines the magnetic concentration and the types of skyrmions realized. It may compromise the phenomenology and the RKKY model for this primary but specific class of skyrmions.
There are a series of studies based on the spin-orbit coupled Kondo-lattice model \cite{Kathyat2020,Mukherjee2021,Kathyat2021,Mukherjee2022,Mukherjee2023},
where the skyrmions are induced by the DM interactions in the double-exchange limit.
They find a Ne\'el-type skyrmion at $t /\lambda\approx 0.55/ 0.45$\cite{Kathyat2021},
which corresponds to $\theta\sim 0.4\pi$ of our model.
\par
The layered cycloidal magnets (cy) hosting N\'eel skyrmions have a different character.
The major DM interactions allowed for these space groups have $\bm D_{ij}$ perpendicular to bonds
and are mostly pointing in-plane, which corresponds to the types of SOC we applied.
Indeed, the SOC of these materials is substantially large\cite{Koelling1977};
the atomic SOC for 3$d$ (Fe, Co), 4$d$ (Ru), and 5$d$ (Ir, Pt) ions are
0.04--0.08 eV, 0.1--0.2 eV, and 0.3--0.6 eV, respectively, and when
combined with the crystal field or the electric potentials coming from strong two-dimensionality,
the electronic states on the clusters or ions have strong anisotropy and mixed spin-angular momentum.
The electronic Hamiltonian in the presence of such effect is represented at the simplest
by our Eq.(\ref{eq:ham})\cite{kurita2011,Witczak2014}.
The value of $\lambda/t$ or $\theta$ is determined
by the strength of SOC and the details of the crystal structures
and can take substantially large values\cite{Nakai2022}.
\par
We thus consider that these materials can be the platform of the skyrmions observed in our model.
Their skyrmions overall have a relatively smaller size than those of the B20-chiral magnets,
which agrees with our results.
In particular, the flake-size skyrmions in our model are the only ones
that can explain the metallic electron skyrmions observed in $C_{nv}$ crystals.
Here, the Mott insulating phases can be partially related to the previously studied spin models
that include a certain amount of quantum fluctuation;
the nano-like skyrmion is found in a semi-classical spin-$S$ model \cite{Maeland2022} that aimed to
explain the 1 nm sized skyrmions found in Fe/Ir layers\cite{Heinze2011}.
The flake-skyrmion can be compared with those observed in the fully quantum spin-1/2 system\cite{Lohani2019}.
However, their magnetic interactions do not conform to the ones naturally obtained from Hubbard models
(see Appendix~\ref{sec:clspin}).
\par
We stress that although previous theories assume the ferromagnetic Heisenberg interactions,
it is elusive for electronic models without SOC.
The superexchange interactions are so far the only natural realization for ferromagnetic exchange,
which shall be the case applied to GaV$_4X_8$ or VOSe$_2$O$_5$ where the V-ions on V$_4X_4$ or VO$_5$
clusters carry spin-1/2 and interact through ligands\cite{Fujima2017,Kezsmarki2015}.
\par
In the Rashba-type SOC Hubbard model on the square lattice before this study,
the double-$q$ spiral phase with noncoplanar spin structures
is found at $0.45\le \theta/\pi \le 0.65$ and $U/t_{\rm eff}\gtrsim 4$\cite{Kawano2023},
which is identified as an antiferromagnetic skyrmion lattice phase, and shall be
relevant to VOSe$_2$O$_5$\cite{Kurumaji2017} and
Sr$X$O$_3$/BaTiO$_3$ hetrostructures\cite{Matsuno2016,Wang2018}.
The future perspective will be to apply a similar analysis to Hubbard models
or other correlated electron models with different fillings or lattices,
which may disclose the landscape of antisymmetric SOC Mott insulators and related SOC metals that host a variety of
spin structures based on spin-split bands.
\section{acknowledgement}
The work is supported by JSPS KAKENHI Grants No. JP21K03440
from the Ministry of Education, Science, Sports and Culture of Japan
and a Grant-in-Aid for Transformative Research Areas "The Natural Laws of Extreme Universe---A New Paradigm for Spacetime and Matter from Quantum Information" (KAKENHI Grant No. 21H05191) from JSPS of Japan.
\appendix
\section{Comparison of SSDMF and SSD-DMRG}\label{sec:dmrg}
We show that SSDMF can reliably capture a long-range correlation
even though we implement a Hartree-Fock level approximation,
which is conventionally considered a primitive approximation
that cannot accommodate any higher order or subtle long-range correlations.
\par
\begin{figure}
  \includegraphics[width=0.45\textwidth]{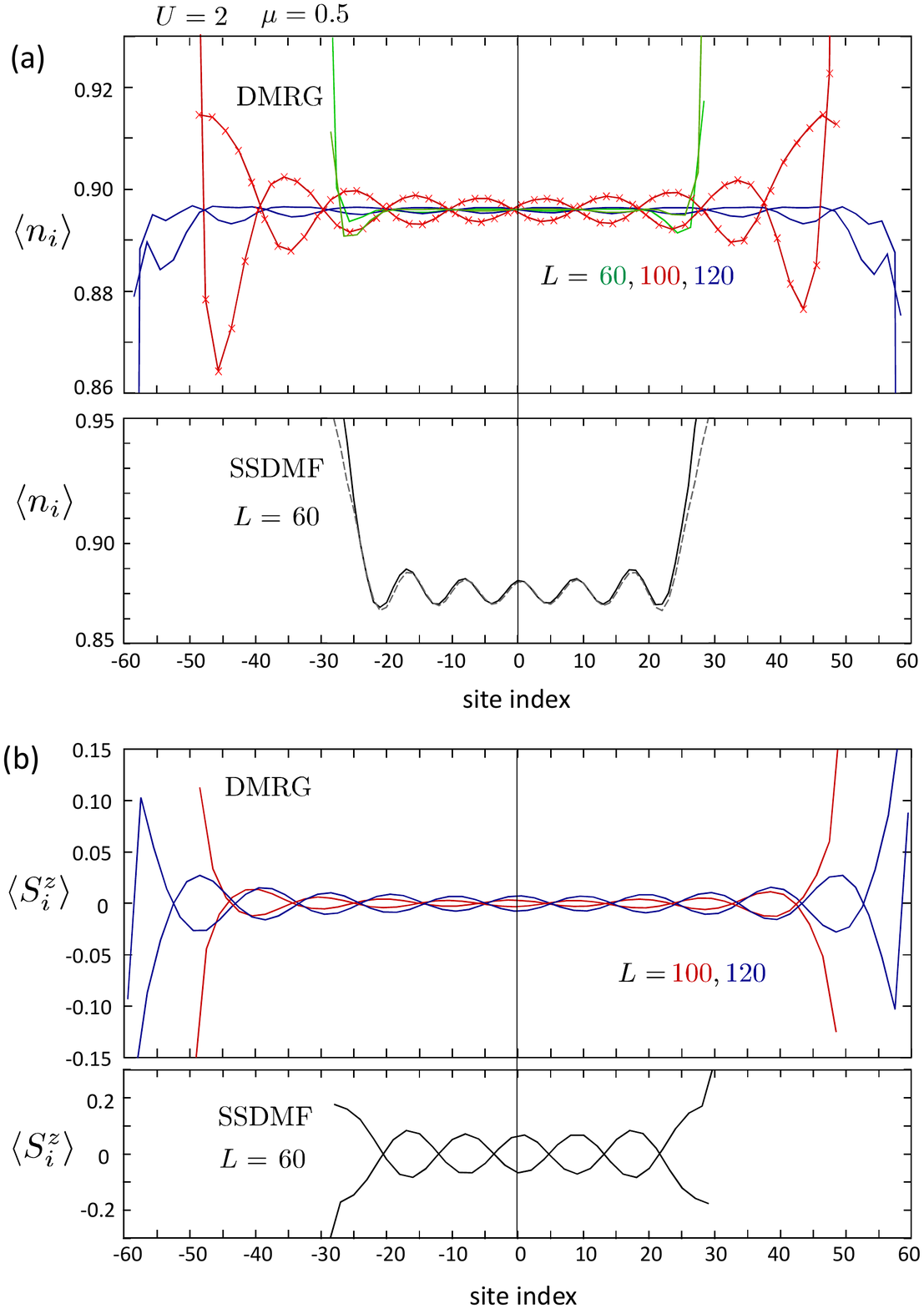}
  \caption{
(a) Charge density $\langle n_i \rangle$ and (b) spin density $\langle S_i^z\rangle$
of a one-dimensional Hubbard model
($\theta=0$ of Eq.(\ref{eq:ham})) with $U=2$ and $\mu=0.5$.
We compare the DMRG combined with SSD using $L=60, 100, 120$,
and SSDMF using $L=60$.
In DMRG we performed more than 10 sweeps and kept the truncation error to less than the order of 10$^{-6}$.
We take the origin of the real space coordinate at the center of
the one-dimensional chain of length $L$.
The deformed Fourier transform of $\langle n_i \rangle$ gives the peak at $Q\sim 0.24\pi$ for $L=60$ and $Q \sim 0.245\pi$ for
$L=100, 120$ in DMRG, consistent with the SSDMF at $Q\sim 0.24\pi$.
Because of the Friedel oscillations and strong correlations, there is an antiferromagnetic correlation
with a period of two lattice sites. We plot two lines that connect data points of even and odd indices, respectively,
to clarify an incommensurate correlation relevant in this state.
For the SSDMF in the lower panel of (a), the broken line is taken from Fig.1(d) in Ref.[\onlinecite{Kawano2022}],
which is reproduced in solid lines we newly calculate.
}
  \label{fS-ssddmrg}
\end{figure}

Figure~\ref{fS-ssddmrg} shows the comparison of DMRG and SSDMF for the
one-dimensional Hubbard model with $U=2$ and $\mu=0.5$,
where for SSDMF, we reproduce the benchmark results shown in Fig. 1 of Ref.[\onlinecite{Kawano2022}].
This parameter gives the state away from half-filling with a charge density of $n_e\sim 0.88$ per site,
where a charge density correlation of $Q \sim 0.24\pi$ appears together with
a spin density correlation $Q \sim 0.88\pi$.
To test the quality of SSDMF,
we apply a DMRG combined with SSD to the same state which takes account of the correlation
effect almost exactly, and is considered the most reliable among all numerical solvers available.
The on-site repulsion induces the misfit of the phase of the charge density of up and down electrons,
and makes the total charge density $\langle n_i\rangle$ and spin density $\langle S_i^z\rangle$
oscillate by $\pi$ in DMRG; this is a typical
Friedel oscillation found for DMRG calculations, an artifact known as a boundary effect\cite{Shibata2011}.
If we connect every two lattice points by lines, we find clear periodicities
that are intrinsic to this quantum state.
\par
In SSD-DMRG, the amplitudes of this oscillation does not depend on the total charge number of the system
because the SSD is not the method to decide the charge density of the total system but to
decide $\mu$ and adopt the charge density at the center part of the system \cite{Hotta2013}.
Indeed, the average charge density (center level of oscillation) and the period almost perfectly
agree between three system sizes, $L=60,100,120$.
By performing a deformed Fourier transform, almost the same wave number, $Q \sim 0.24\pi-0.245\pi$,
is captured for DMRG as those of SSDMF.
\par
We note that for a usual periodic boundary, the one-body on-site quantities do not show any periodicity due to
translational symmetry and the two-point correlation stores the information of the phase.
However, in SSD systems the translational symmetry is broken, and the on-site quantities
are modulated by the wave number that captures the dominant correlation of the state.

%
\section{Several magnetic ground states in the phase diagrams}\label{sec:appmagstates}
\begin{figure}
  \includegraphics[width=0.45\textwidth]{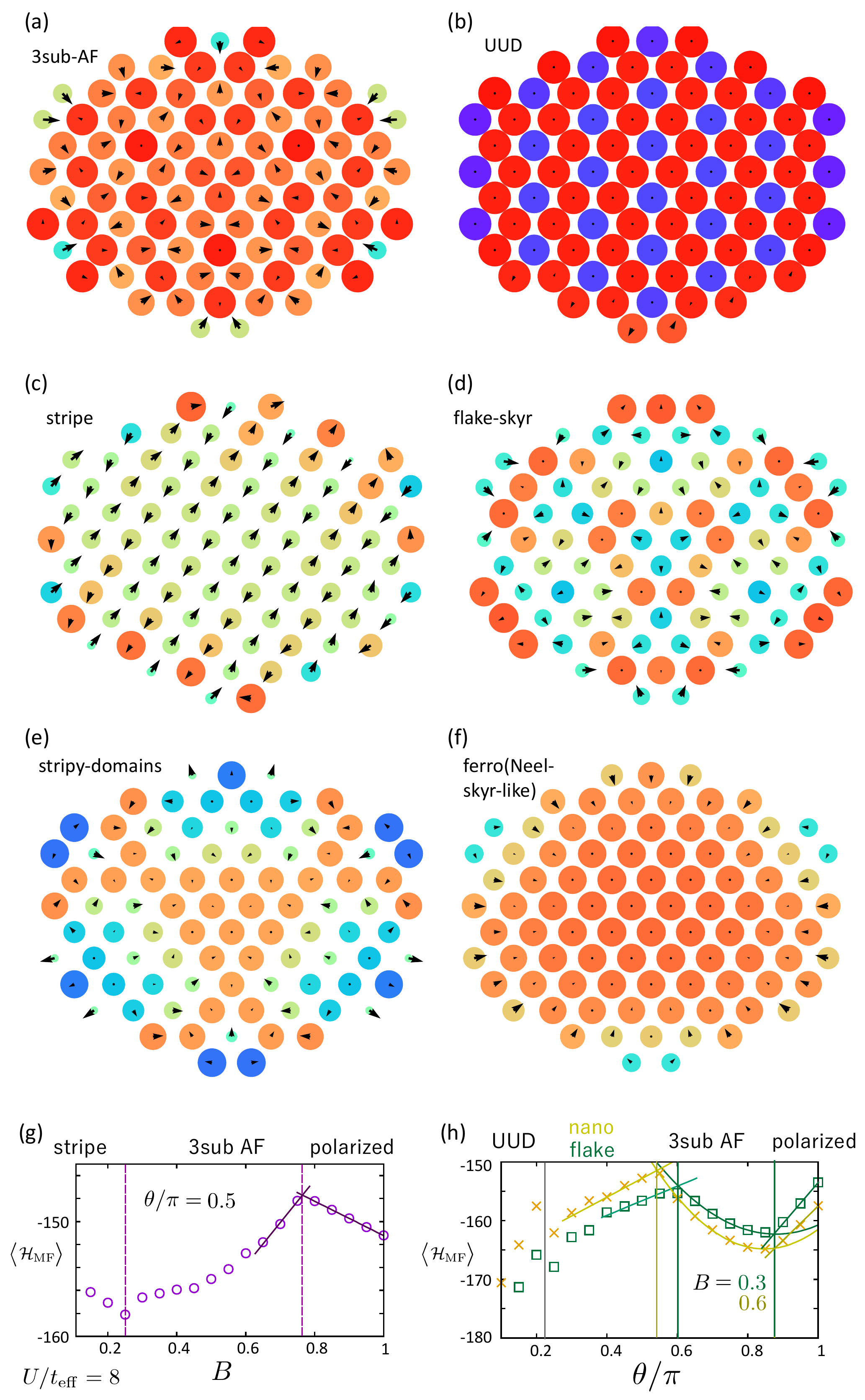}
  \caption{Real space spin textures of several phases found in Figs.~1(b) and~1(c) in the main text.
(a)-(c) $U/t_{\rm eff}=8$ states: 3 sub-AF $(B,\theta/\pi)=(0.8,0.4)$,
UUD $(0.5,0)$, and
stripe $(0.1,0.55)$.
(d)-(f) $U/t_{\rm eff}=5$ states: flake-skyr$(B,\theta/\pi)=(0.2,0.35)$,
stripy-domains $(0.2,0.65)$, and
N\'eel-like (ferro) $(0.7,0.55)$.
The colors of the circles are the density plots of $\langle S_i^z\rangle$
and arrows denote the in-plane element as vectors ($\langle S_i^x\rangle,\langle S_i^y\rangle$).
Energy of the mean-field solution for fixed (g) $\theta/\pi=0.5$ with varying $B$
and (h) $B=0.3,0.6$ with varying $\theta$, where the corresponding phases are shown.
}
  \label{fS-states}
\end{figure}

We present in Fig.~\ref{fS-states} the spin configuration of each phase in
Figs.~1(a) and~ 1(b) not referred to in detail in the main text.
At $U/t_{\rm eff}=8$ the 3 sub-sky and 3 sub-AF (120$^\circ$-like) magnetic states appear
at large $B$ whose moments are relatively large (Figs.~\ref{f3sub}(a), ~\ref{fS-states}(a)).
\par
The UUD state at $\theta=0$ given in Fig.~\ref{fS-states}(b)
is the typical phase observed in quantum Heisenberg or Ising models on a triangular lattice in the magnetic field,
often forming plateaus.
The stripe phase emerges when the system approaches the ferromagnetic phase at a small $B$
which is shown in Fig.~\ref{fS-states}(c).
\par
The states at $U/t_{\rm eff}=5$ are mostly metallic,
so the magnetic moments are much smaller than in the above cases.
The flake-skyr phase (Fig.~\ref{fS-states}(d)) is very stable, which is the one discussed mainly in the main text.
The stripy-domains (Fig.~\ref{fS-states}(e)) that have similarity with N\'eel-type skyrmion.
\par
The one found in the smaller $\theta$ region (shaded) of the ferromagnetic phase
in Fig.~1(b,c) in the main text has spin structure shown in Fig.~\ref{fS-states}(f),
and is classified as a polarized ferromagnet,
although there is a small but finite modulation of spin moments
similar to N\'eel-type skyr.
We have carefully examined the energy differences of this SSDMF solution
with those of the fully polarized ferromagnet by
choosing the proper cluster with periodic boundary condition
and recalculating the solution for the uniform (non-SSD) Hamiltonian,
finding that the fully polarized ferromagnet slightly overwhelms
the N\'eel-like skyrmions.
The parameter range where this kind of solution appears is shaded in the phase diagram.
We expect that the N\'eel-like skyrmions may appear at smaller $\theta$
part of this region at finite temperature as in the previously reported theories.
%
\begin{figure}
  \includegraphics[width=0.45\textwidth]{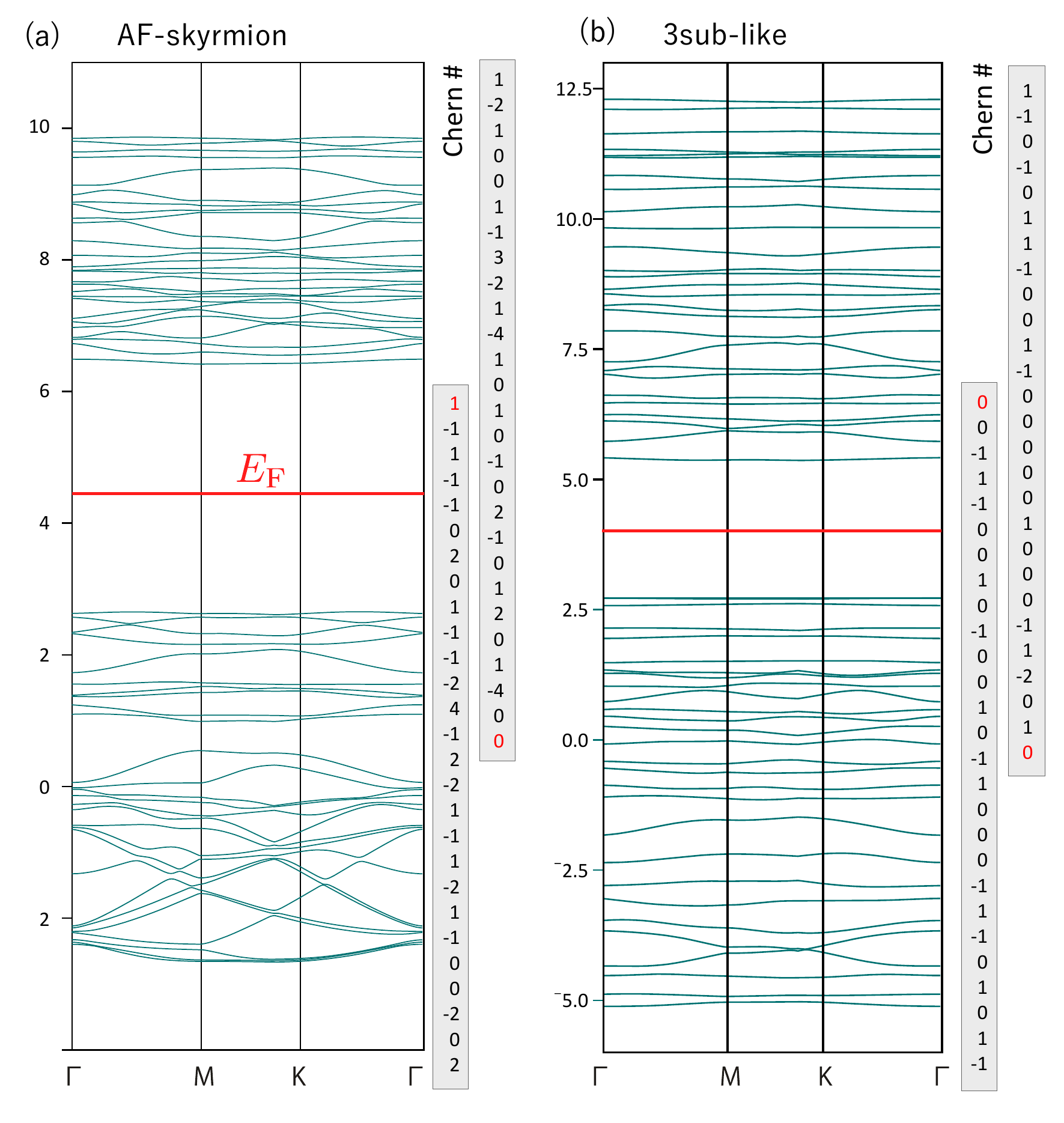}
  \caption{Energy bands of insulating states at $U/t_{\rm eff}=8$
with (a) AF-skyrmions $(B/t_{\rm eff},\theta/\pi)=(0.1,0.5)$ shown in Fig.~1(f) in the main text
and (b) UUD 3 sub-like state at $(0.45,0.15)$.
Chern numbers of bands are shown on the r.h.s. In (a) it has a nonzero value at the valence band.
}
  \label{fS-ins}
\end{figure}
%
\begin{figure}
  \includegraphics[width=0.45\textwidth]{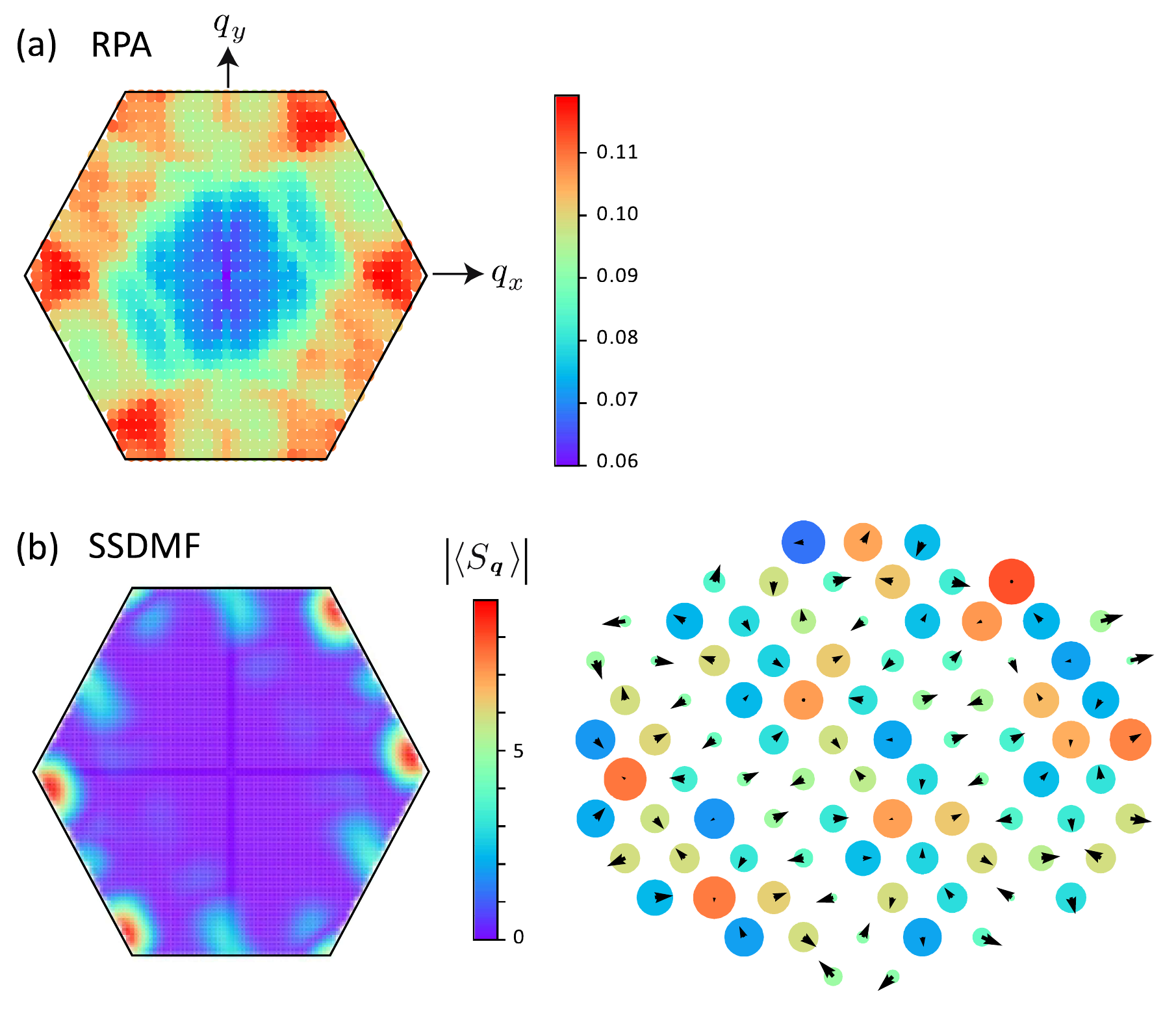}
  \caption{
(a) Density plot of the maximum eigenvalue of the bare magnetic susceptibility $\lambda_3(\bm{q})$ as a
function of $\bm{q}$, obtained for $\theta=0.3\pi$.
(b) SSDMF solution for $\theta=0.3\pi$ and $U=5$.
Left panel is the density plot of the amplitude of spin moment $|\langle \bm S_{\bm q}\rangle|$
in the Brillouin zone, and the right panel shows
the corresponding spin structure in real space.
}
  \label{fS-RPA}
\end{figure}
\par
The phase boundaries are obtained by examining how the energy $\langle {\cal H}_{\rm MF}\rangle$
varies with $\theta$ and $B$, whose examples are shown in Figs.~\ref{fS-states}(g) and \ref{fS-states}(h).
There is a distinct change in the functional form of energy,
which can be analyzed to determine the phase diagram
together with the careful examination of the real space and $k$-space magnetic structures.
\par
We next show the energy band structures of insulating AF-skyrmion and UUD 3 sub-like states
in Fig.~Sref{fS-ins}.
They are realized in the $U/t_{\rm eff}=8$ phase diagram Fig.~1(b) at
$(B/t_{\rm eff},\theta/\pi)=(0.1,0.5)$ and $(0.45,0.15)$, respectively.
The energy dispersion is suppressed compared to Fig.~3(b)
and a large charge gap of $\sim t_{\rm eff}$ opens.
The Chern numbers are shown on the r.h.s. of the energy bands.
Since the real-space magnetic textures
(AF-skyrmion is shown in Fig.~1(f) and 3 sub-like state similar to 3 sub-cluster in Fig.~\ref{fS-states}(a))
do not seem to differ too much from
the metallic flake-skyrmion shown in Fig.~3(a) (main text),
they have a nonzero winding number,
while Chern numbers near the Fermi level are both zero for the latter
and the valence band is 1 for the former.
\section{Random phase approximations}\label{sec:rpa}
To support the reliability of the spin structure found
in the phase diagram of Fig. 1(c) we perform a random phase approximation (RPA) to elucidate the instability
toward the ordered phases.
We start from a magnetic susceptibility at $U=0$ given by
\begin{equation}
  \chi_0^{\mu,\nu}(\bm{q}) =
  -\frac{1}{N}\sum_{\bm{k},n,m}
  s^{\mu}_{n,m}(\bm{k},\bm{k}+\bm{q})
  s^{\nu}_{m,n}(\bm{k}+\bm{q},\bm{k})
  F_{n,m}(\bm{k},\bm{k}+\bm{q})
\end{equation}
where $\mu,\nu=x,y,z$.
Because the model has a SOC, different spin orientations are mixed in forming energy bands,
and the corresponding $\chi_0(\bm{q})$ is given in the $3\times 3$ matrix form,
whose elements are evaluated using
\begin{gather}
  F_{n,m}(\bm{k}, \bm{k}+\bm{q})
  =\frac{f(\varepsilon_n(\bm{k}))-f(\varepsilon_m(\bm{k}+\bm{q}))}
  {\varepsilon_n(\bm{k})-\varepsilon_m(\bm{k}+\bm{q})+i\hbar\delta},\\
  s^{\mu}_{n,m}(\bm{k}_1,\bm{k}_2)=
  \bm{u}^\dagger_n(\bm{k}_1)
  \qty(\frac{\sigma^{\mu}}{2})
  \bm{u}_m(\bm{k}_2).
\end{gather}
Here, $f(\varepsilon)=1/(\exp(\varepsilon/k_{\rm B}T)+1)$ is the Fermi distribution function,
$\delta$ is an infinitesimal positive number, and $\varepsilon_m(\bm{k})$ and $\bm{u}_m(\bm{k})$ are
the eigenvalue and corresponding eigenvector ($m=\pm$) of the Bloch Hamiltonian given by
\begin{gather}
  \varepsilon_{\pm}(\bm{k})=-2t{\rm Re}w(\bm{k})\pm 2\lambda |{\rm Im}\bm{v}(\bm{k})|,\\
  \bm{u}_{\pm}(\bm{k})=\frac{1}{\sqrt{2}}
  \begin{pmatrix}
   1 \\ \pm e^{i\phi}
  \end{pmatrix},\\
  w(\bm{k})=\sum_{j=1,2,3}e^{-i\bm{k}\cdot \bm{e}_j},\\
  \bm{v}(\bm{k})=\sum_{j=1,2,3}e^{-i\bm{k}\cdot \bm{e}_j}\bm{n}_{j},
\end{gather}
where $\bm{e}_{j}$ denotes the primitive translation vectors for the triangular lattice,
$\bm{n}_{j}$ is the direction of the rotation axis for SOC on the edge corresponding to $\bm{e}_{j}$, and $\phi$ is defined by ${\rm Im}\bm{v}=|{\rm Im}\bm{v}|{}^t(\cos\phi,\sin\phi)$.
The RPA susceptibility is given in the $3\times 3$ form as
\begin{equation}
\chi_{\rm RPA}(\bm{q}) = [I_{3\times3}-2U\chi_0(\bm{q})]^{-1}\chi_0(\bm{q}).
\end{equation}
Once the eigenvalues of $\chi_0(\bm{q})$ are obtained as $\lambda_j$ ($j=1,2,3$) ,
the one with the largest amplitude, $\lambda_3(\bm{q})$,
that satisfies $1-2U\max_{\bm{q}}\lambda_3(\bm{q})=0$,
will determine the phase transition point $U=U_c$,
and we find the possible ordering wave vector $\bm{Q}={\rm argmax}_{\bm{q}}\lambda_3(\bm{q})$, at $U>U_c$.
\par
Figure~\ref{fS-RPA}(a) shows the density plot of $\lambda_3(\bm{q})$ for $\theta = 0.3\pi$
and from its highest peak value, the critical point is estimated as $U_{\rm c}=4.19$.
We find very good agreement between the ordering wave vector obtained by RPA and that obtained by
SSDMF for $\theta=0.3\pi$ and $B=0$, shown in Fig.~\ref{fS-RPA} (b).
Therefore, SSDMF safely captures the incommensurate wave number,
which is difficult for conventional numerical methods based on the periodic boundary condition.
\par
In general, the Hartree-Fock or RPA approaches may overestimate the stability of ordered phases
and $U_c$ shall be increased when the correlation effects are taken into account.
However, in the previous paper for the square lattice Hubbard model with SOC,
DMET has supported the emergent magnetic order at $U=2$, consistently with SSDMF\cite{Kawano2023}.
This may resolve the concern that the incommensurate magnetic orders at the mean-field level
may be wiped out by the fluctuation when we introduce higher-order correlation effects.
Considering that the Mott transition in triangular lattice Hubbard model without SOC but with anisotropy occurs
for most of the results with quantum many-body numerical solvers at around $U_c\sim 4-6$ \cite{Hotta2012crystal},
it is natural to expect that our phase diagram at $U\geq 5$ accommodates magnetically ordered phases, which may
not necessarily an insulator.
\section{Comparison with several classical spin models}\label{sec:clspin}
To understand how the types of magnetic interactions used in the previous studies influence the
ground state phase diagram,
we consider the analogues of second order Hamiltonian ${\mathcal H}^{(2)}$ in Eq.(7) in the main text,
taking $\bm S_i$ as classical vector of size-1.
In the main text, the results for Eq.(7) (${\mathcal H}^{(2)}$)
with $\bm n$ placed in the $xy$-plane perpendicular to the $ij$-bonds
is shown in Fig.~4(a) where the parameters $J$, $D$ and $K$ are the functions of $\theta$.
This result was compared with its analogs naturally derived by setting $K=0$ in Fig.~4(b)
or by taking $\bm n \parallel z$-axis in Fig.~4(c).
\par
\begin{figure}
  \includegraphics[width=0.5\textwidth]{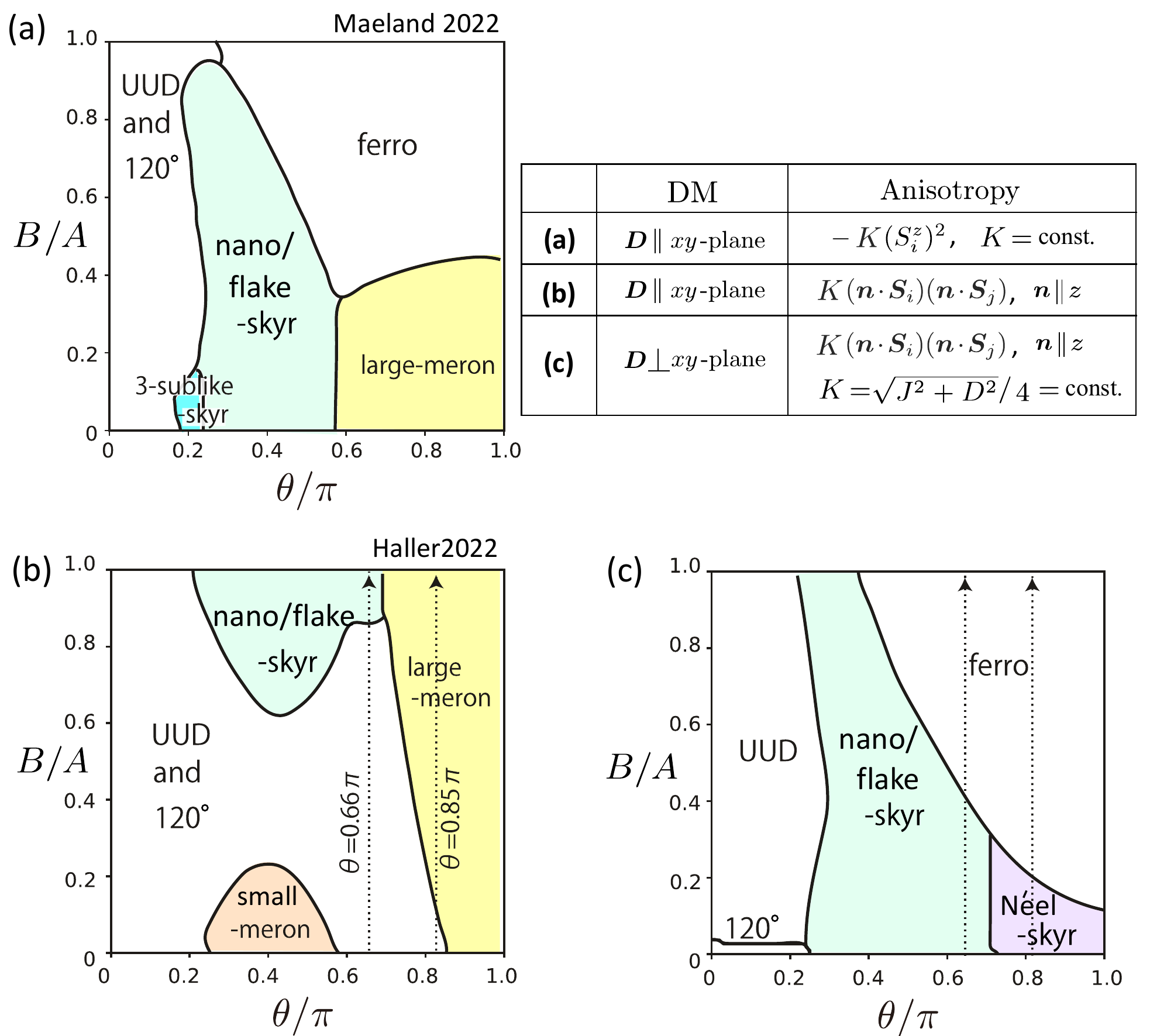}
  \caption{Ground state phase diagram of the classical spin model with different types of anisotropies.
  The interactions $J=\cos\theta, D=\sin\theta$ are set as the same as Fig.~4(main text),
  with $B$ given in unit of $A=4t_{\rm eff}^{2}/U=0.4$.
  For panel (a) we take $-K (S_i^z)^2$ with constant $K$
  given as $(4/U)(K_m/\sqrt{D_m^2+1})$ with $K_m\equiv K/J=0.518$ and $D_m= D/J= 2.16$ following Ref.[\onlinecite{Maeland2022}].
 In panels (b) and (c) we considered the XXZ type anisotropies by setting $\bm n\parallel z$
  and taking $K=1-\cos\theta$ for (b) and $K=\sqrt{J^2+D^2}/4$ being constant for (c),
 to compare with Ref.[\onlinecite{Haller2022}] whose $J=-2D$ and $J=-0.5D$ cases are
  shown in broken lines $\theta=0.66\pi$ and $0.85\pi$, respectively.
}
  \label{fS-other}
\end{figure}

However, the types of interactions or anisotropies adopted in previous studies are different;
in Ref.[\onlinecite{Maeland2022}] they took $\bm D\perp$ bond in the $xy$-plane as in our case
but considered easy-axis anisotropy which is valid only for $S>1$ as
$-K\sum_i (S_i^z)^2$.
To see this effect we calculate in Fig.~\ref{fS-other}(a) the classical ground state phase diagram
by setting $K$ as constant, while varying $D$ and $J$ as functions of $\theta$.
The constant values, $K=(4/U)(K_m/\sqrt{D_m^2+1})$ with $K_m=0.518$ and $D_m=2.16$, are set to
directly compare with the case of Ref.[\onlinecite{Maeland2022}].
Their SkX1 phase has a similar structure to our nano/flake-skyrmion and is consistent with the
stable skyr region in the phase diagram.
\par
Regarding Ref.[\onlinecite{Haller2022}],
they used $\bm D\perp$ bond in the $xy$-plane but for spin anisotropies
they set $\bm n \parallel z$
(corresponding to the XXZ-model, taking the spin anisotropic exchange interaction),
which does not conform to the symmetries provided by the SOC in the Hubbard model.
The phase diagram obtained by varying $J, D, K$ in the same manner as our model
but taking $\bm n\parallel z$ for the anisotropy term to follow Ref.[\onlinecite{Haller2022}]
is shown in Fig.~\ref{fS-other}(b).
The ones taking $K=\sqrt{J^2+D^2}/4$ as constant about $\theta$ is also shown
in Fig.~\ref{fS-other}(c).
Broken lines are $\theta=0.66\pi$ and $0.85\pi$ studied in Ref.[\onlinecite{Haller2022}]
as $J=-2D$ and $J=-0.5D$, respectively.
If we interpret our nano/flake-skyr as their SKX$_{3a}$,
the results of ours and theirs are slightly inconsistent in that the nano/flake-skyr appear
only in $J=-2D$ but not in $J=-0.5D$, while they argue that both cases have SKX$_{3a}$.
In that context, our phase diagram is natural
since the $\theta=0.85\pi$, where $|J|>|D|$ holds, favors larger size skyrmion.
\par
Although these works stress the quantum mechanical magnetic fluctuation as being key to
realize small skyrmions, the systematic comparison given here suggests that the relationships between magnetic interactions
governed by SOC seems to play a more important role.

\section{Effect of ring exchange in fourth-order perturbation}\label{sec:4th}
In Fig.~4 in the main text, we examined ${\mathcal H}^{(2)}$ including interactions
originating from the second-order perturbation terms.
Since the role of higher order terms is being discussed recently\cite{Hayami2017},
we additionally consider the fourth-order spin-dependent ring exchange term.
By considering the SOC hopping term $\lambda$, the ring exchange term is given in the form,
\begin{equation}
  \mathcal{H}_{\rm ring}^{(4)} = \sum_{abcd} \sum_{j=0}^4
\hat R_j [\bm{S}_a,\bm{S}_b,\bm{S}_c,\bm{S}_d] \cos^j\frac{\theta}{2} \sin^{4-j}\frac{\theta}{2},
  \label{four-body}
\end{equation}
where $\hat R_j$ is a four-body operator about the four spins $a,b,c,d$ on a unit plaquette consisting of two triangles,
and its form depends on $j$.
At $\theta=0$ this term reduces to the normal ring exchange term of the Hubbard model at half-filling
\cite{Takahashi1977},
which takes the form,
\begin{equation}
J_4 \sum_{ abcd} \big( (s_a\cdot s_b)(s_c \cdot s_d)+(s_a\cdot s_d)(s_b \cdot s_c)
-(s_a\cdot s_c)(s_b \cdot s_d)\big).
\end{equation}
This conventional ring exchange term is known to give a substantial influence on the magnetism
particularly when $U/t_{\rm eff}$ becomes small near the metal-insulator phase boundary.
\par
\begin{figure}
  \includegraphics[width=0.4\textwidth]{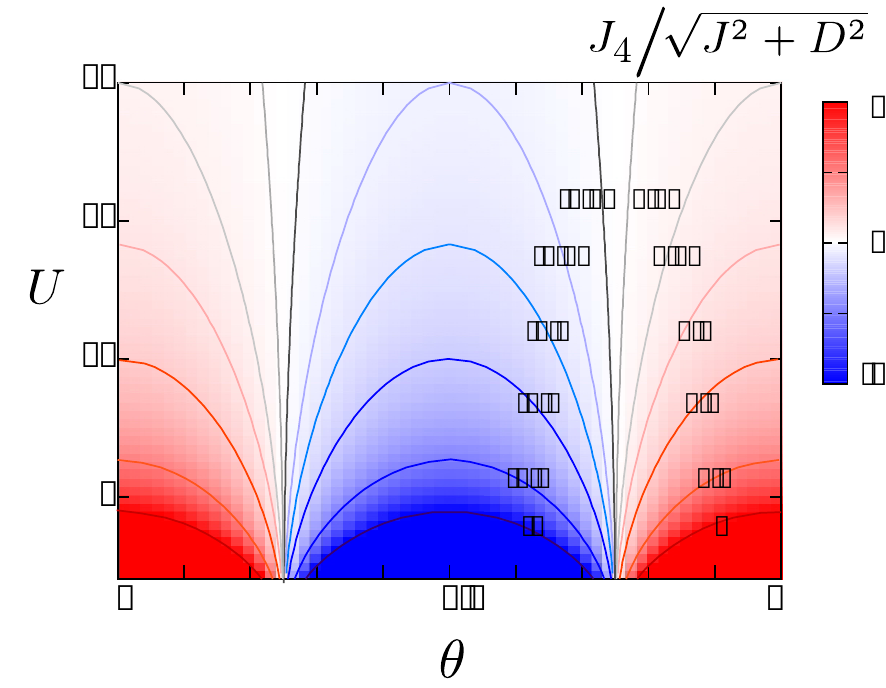}
  \caption{Density (contour) plot of $J_4/\sqrt{D^2+J^2}$ on the plane of $\theta$ and $U$,
  where we set $t_{\rm eff}=1$ and compared the ratio of the major fourth-order ring exchange interaction
  and the 2nd order magnetic interaction $\sqrt{D^2+J^2}=A=4t_{\rm eff}^2/U$.
}
  \label{fS-ring}
\end{figure}

At $\theta\ne 0$, the coupling constant is given as $J_4(\theta)=80t^4 \cos(2\theta)/U^3$.
To see how much contribution this term gives compared to the second-order terms,
we show in Fig.~\ref{fS-ring} the contour plot of $J_4(\theta)/\sqrt{D^2+J^2}$
as a function of $\theta$ and $U$ where we set $t_{\rm eff}=1$.
At around $U=8$, $|J_4|/\sqrt{D^2+J^2} \lesssim 0.2$,
showing that the effect of ring exchange interaction is small for the insulating phase.
However, when $U=5$ the contribution from the fourth-order terms is similar
to the second-order terms, while since the system is in the metallic region at $\theta \lesssim 0.6$,
these magnetic interactions no longer make sense.
We may interpret the enhancement of $J_4$ as having
an itinerancy in the magnetism, which increases the quantum fluctuation effect.
Then, based on this analysis, the comparison between the phase diagrams
in Figs.~1(b) and 1(c) (main text)
can be made; although the magnetic phases become less distinct for the smaller $U$,
the basic features of magnetism do not differ much.
Therefore, we can conclude that the fourth-order terms may play only a secondary role
for the stabilization of the skyrmion phases.

\section{Magnetic-field dependence of ground state band structures and Chern numbers}\label{sec:Chern}
\begin{figure*}
  \includegraphics[width=0.85\textwidth]{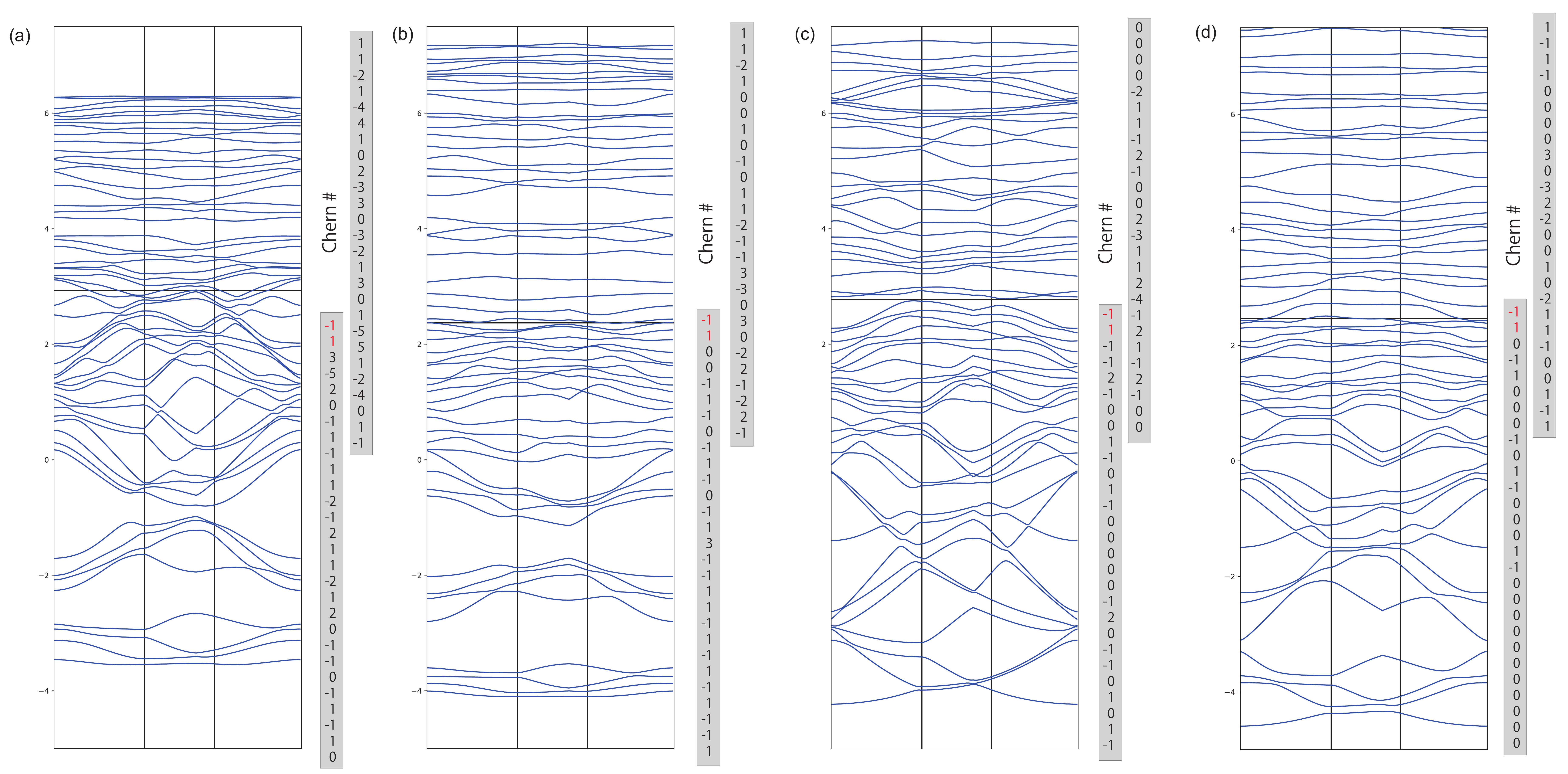}
  \caption{Energy bands of flake skyrmions at $U/t_{\rm eff}=5$ with (a) $(B/t_{\rm eff},\theta/\pi)=(0.2,0.35)$. (b) (0.35,0.35).  (c) (0.75,0.35). (d) (1,0.35).
}
  \label{fS-Chern}
\end{figure*}

We examine how the band structure of the flake skyrmion at $U/t_{\rm eff}=5$ (see Fig.~\ref{f2})
evolves with the field.
Here, we fix $\theta/\pi=0.35$ and plot the energy band and the Chern numbers for several choices of $B$ in Fig.~\ref{fS-Chern}.
As the band structure varies gradually, the Chern numbers change throughout the band, while
the two bands at the Fermi level seem to have $\pm 1$.
\bibliographystyle{naturemag}
\bibliography{ref}

\begin{thebibliography}{10}
\expandafter\ifx\csname url\endcsname\relax
  \def\url#1{\texttt{#1}}\fi
\expandafter\ifx\csname urlprefix\endcsname\relax\def\urlprefix{URL }\fi
\providecommand{\bibinfo}[2]{#2}
\providecommand{\eprint}[2][]{\url{#2}}

\bibitem{Skyrme1961}
\bibinfo{author}{Skyrme, T.~H.}
\newblock \bibinfo{title}{{A non-linear field theory}}.
\newblock \emph{\bibinfo{journal}{Proc. R. Soc. Lond. A}}
  \textbf{\bibinfo{volume}{260}}, \bibinfo{pages}{127–--138}
  (\bibinfo{year}{1961}).

\bibitem{Sondhi1993}
\bibinfo{author}{Sondhi, S.~L.}, \bibinfo{author}{Karlhede, A.},
  \bibinfo{author}{Kivelson, S.~A.} \& \bibinfo{author}{Rezayi, E.~H.}
\newblock \bibinfo{title}{{Skyrmions and the crossover from the integer to
  fractional quantum Hall effect at small {Zeeman} energies}}.
\newblock \emph{\bibinfo{journal}{Phys. Rev. B}} \textbf{\bibinfo{volume}{47}},
  \bibinfo{pages}{16419--16426} (\bibinfo{year}{1993}).

\bibitem{Tycko1995}
\bibinfo{author}{Tycko, R.}, \bibinfo{author}{Barrett, S.~E.},
  \bibinfo{author}{Dabbagh, G.}, \bibinfo{author}{Pfeiffer, L.~N.} \&
  \bibinfo{author}{West, K.~W.}
\newblock \bibinfo{title}{{Electronic states in gallium arsenide quantum wells
  probed by optically pumped {NMR}}}.
\newblock \emph{\bibinfo{journal}{Science}} \textbf{\bibinfo{volume}{268}},
  \bibinfo{pages}{1460--1463} (\bibinfo{year}{1995}).

\bibitem{Bogdanov1989-2}
\bibinfo{author}{Bogdanov, A.}
\newblock \bibinfo{title}{{Thermodynamically stable vortexes in magnetically
  ordered crystals-mixed state of magnetics}}.
\newblock \emph{\bibinfo{journal}{Zh. Eksp. Teor. Fiz.}}
  \textbf{\bibinfo{volume}{95}}, \bibinfo{pages}{178--182}
  (\bibinfo{year}{1989}).

\bibitem{Muhlbauer2009}
\bibinfo{author}{Muhlbauer, S.} \emph{et~al.}
\newblock \bibinfo{title}{{Skyrmion Lattice in a Chiral Magnet}}.
\newblock \emph{\bibinfo{journal}{Science}} \textbf{\bibinfo{volume}{323}},
  \bibinfo{pages}{915--919} (\bibinfo{year}{2009}).

\bibitem{Neubauer2009}
\bibinfo{author}{Neubauer, A.} \emph{et~al.}
\newblock \bibinfo{title}{{Topological Hall Effect in the $A$ Phase of MnSi}}.
\newblock \emph{\bibinfo{journal}{Phys. Rev. Lett.}}
  \textbf{\bibinfo{volume}{102}}, \bibinfo{pages}{186602}
  (\bibinfo{year}{2009}).

\bibitem{Yu2010}
\bibinfo{author}{Yu, X.~Z.} \emph{et~al.}
\newblock \bibinfo{title}{{Real-space observation of a two-dimensional skyrmion
  crystal}}.
\newblock \emph{\bibinfo{journal}{Nature}} \textbf{\bibinfo{volume}{465}},
  \bibinfo{pages}{901--904} (\bibinfo{year}{2010}).

\bibitem{Yu2011}
\bibinfo{author}{Yu, X.~Z.} \emph{et~al.}
\newblock \bibinfo{title}{{Near room-temperature formation of a skyrmion
  crystal in thin-films of the helimagnet FeGe}}.
\newblock \emph{\bibinfo{journal}{Nature Materials}}
  \textbf{\bibinfo{volume}{10}}, \bibinfo{pages}{106--109}
  (\bibinfo{year}{2011}).

\bibitem{Seki2012}
\bibinfo{author}{Seki, S.}, \bibinfo{author}{Yu, X.~Z.},
  \bibinfo{author}{Ishiwata, S.} \& \bibinfo{author}{Tokura, Y.}
\newblock \bibinfo{title}{{Observation of Skyrmions in a Multiferroic
  Material}}.
\newblock \emph{\bibinfo{journal}{Science}} \textbf{\bibinfo{volume}{336}},
  \bibinfo{pages}{198--201} (\bibinfo{year}{2012}).

\bibitem{Tokunaga2015}
\bibinfo{author}{Tokunaga, Y.} \emph{et~al.}
\newblock \bibinfo{title}{{A new class of chiral materials hosting magnetic
  skyrmions beyond room temperature}}.
\newblock \emph{\bibinfo{journal}{Nature Communications}}
  \textbf{\bibinfo{volume}{6}}, \bibinfo{pages}{7638} (\bibinfo{year}{2015}).

\bibitem{Nagaosa2013}
\bibinfo{author}{Nagaosa, N.} \& \bibinfo{author}{Tokura, Y.}
\newblock \bibinfo{title}{{Topological properties and dynamics of magnetic
  skyrmions}}.
\newblock \emph{\bibinfo{journal}{Nature Nanotechnology}}
  \textbf{\bibinfo{volume}{8}}, \bibinfo{pages}{899--911}
  (\bibinfo{year}{2013}).

\bibitem{Kezsmarki2015}
\bibinfo{author}{K{\'e}zsm{\'a}rki, I.} \emph{et~al.}
\newblock \bibinfo{title}{{N{\'e}el-type skyrmion lattice with confined
  orientation in the polar magnetic semiconductor {Ga}{V}$_4${S}$_8$}}.
\newblock \emph{\bibinfo{journal}{Nature Materials}}
  \textbf{\bibinfo{volume}{14}}, \bibinfo{pages}{1116--1122}
  (\bibinfo{year}{2015}).

\bibitem{Fujima2017}
\bibinfo{author}{Fujima, Y.}, \bibinfo{author}{Abe, N.},
  \bibinfo{author}{Tokunaga, Y.} \& \bibinfo{author}{Arima, T.}
\newblock \bibinfo{title}{{Thermodynamically stable skyrmion lattice at low
  temperatures in a bulk crystal of lacunar spinel {GaV}$_{4}${Se}$_{8}$}}.
\newblock \emph{\bibinfo{journal}{Phys. Rev. B}} \textbf{\bibinfo{volume}{95}},
  \bibinfo{pages}{180410} (\bibinfo{year}{2017}).

\bibitem{Butykai2017}
\bibinfo{author}{Butykai, A.} \emph{et~al.}
\newblock \bibinfo{title}{{Characteristics of ferroelectric-ferroelastic
  domains in {N}{\'e}el-type skyrmion host {GaV$_4$S$_8$}}}.
\newblock \emph{\bibinfo{journal}{Scientific Reports}}
  \textbf{\bibinfo{volume}{7}}, \bibinfo{pages}{44663} (\bibinfo{year}{2017}).

\bibitem{Kurumaji2017}
\bibinfo{author}{Kurumaji, T.} \emph{et~al.}
\newblock \bibinfo{title}{{{N}{\'e}el-Type Skyrmion Lattice in the Tetragonal
  Polar Magnet ${\mathrm{VOSe}}_{2}{\mathrm{O}}_{5}$}}.
\newblock \emph{\bibinfo{journal}{Phys. Rev. Lett.}}
  \textbf{\bibinfo{volume}{119}}, \bibinfo{pages}{237201}
  (\bibinfo{year}{2017}).

\bibitem{Kurumaji2019}
\bibinfo{author}{Kurumaji, T.} \emph{et~al.}
\newblock \bibinfo{title}{{Skyrmion lattice with a giant topological {Hall}
  effect in a frustrated triangular-lattice magnet}}.
\newblock \emph{\bibinfo{journal}{Science}} \textbf{\bibinfo{volume}{365}},
  \bibinfo{pages}{914--918} (\bibinfo{year}{2019}).

\bibitem{Hirschberger2019}
\bibinfo{author}{Hirschberger, M.} \emph{et~al.}
\newblock \bibinfo{title}{{Skyrmion phase and competing magnetic orders on a
  breathing kagom{\'e} lattice}}.
\newblock \emph{\bibinfo{journal}{Nature Commun.}}
  \textbf{\bibinfo{volume}{10}}, \bibinfo{pages}{5831} (\bibinfo{year}{2019}).

\bibitem{Okubo2012}
\bibinfo{author}{Okubo, T.}, \bibinfo{author}{Chung, S.} \&
  \bibinfo{author}{Kawamura, H.}
\newblock \bibinfo{title}{{Multiple-$q$ States and the Skyrmion Lattice of the
  Triangular-Lattice Heisenberg Antiferromagnet under Magnetic Fields}}.
\newblock \emph{\bibinfo{journal}{Phys. Rev. Lett.}}
  \textbf{\bibinfo{volume}{108}}, \bibinfo{pages}{017206}
  (\bibinfo{year}{2012}).

\bibitem{Hayami2017}
\bibinfo{author}{Hayami, S.}, \bibinfo{author}{Ozawa, R.} \&
  \bibinfo{author}{Motome, Y.}
\newblock \bibinfo{title}{{Effective bilinear-biquadratic model for noncoplanar
  ordering in itinerant magnets}}.
\newblock \emph{\bibinfo{journal}{Phys. Rev. B}} \textbf{\bibinfo{volume}{95}},
  \bibinfo{pages}{224424} (\bibinfo{year}{2017}).

\bibitem{Wang2020}
\bibinfo{author}{Wang, Z.}, \bibinfo{author}{Su, Y.}, \bibinfo{author}{Lin,
  S.-Z.} \& \bibinfo{author}{Batista, C.~D.}
\newblock \bibinfo{title}{{Skyrmion Crystal from {RKKY} Interaction Mediated by
  {2D} Electron Gas}}.
\newblock \emph{\bibinfo{journal}{Phys. Rev. Lett.}}
  \textbf{\bibinfo{volume}{124}}, \bibinfo{pages}{207201}
  (\bibinfo{year}{2020}).

\bibitem{Mitsumoto2021}
\bibinfo{author}{Mitsumoto, K.} \& \bibinfo{author}{Kawamura, H.}
\newblock \bibinfo{title}{{Replica symmetry breaking in the {RKKY}
  skyrmion-crystal system}}.
\newblock \emph{\bibinfo{journal}{Phys. Rev. B}}
  \textbf{\bibinfo{volume}{104}}, \bibinfo{pages}{184432}
  (\bibinfo{year}{2021}).

\bibitem{Yi2009}
\bibinfo{author}{Yi, S.~D.}, \bibinfo{author}{Onoda, S.},
  \bibinfo{author}{Nagaosa, N.} \& \bibinfo{author}{Han, J.~H.}
\newblock \bibinfo{title}{{Skyrmions and anomalous Hall effect in a
  {Dzyaloshinskii}-{Moriya} spiral magnet}}.
\newblock \emph{\bibinfo{journal}{Phys. Rev. B}} \textbf{\bibinfo{volume}{80}},
  \bibinfo{pages}{054416} (\bibinfo{year}{2009}).

\bibitem{Buhrandt2013}
\bibinfo{author}{Buhrandt, S.} \& \bibinfo{author}{Fritz, L.}
\newblock \bibinfo{title}{{Skyrmion lattice phase in three-dimensional chiral
  magnets from {Monte} {Carlo} simulations}}.
\newblock \emph{\bibinfo{journal}{Phys. Rev. B}} \textbf{\bibinfo{volume}{88}},
  \bibinfo{pages}{195137} (\bibinfo{year}{2013}).

\bibitem{Witczak2014}
\bibinfo{author}{Witczak-Krempa, W.}, \bibinfo{author}{Chen, G.},
  \bibinfo{author}{Kim, Y.~B.} \& \bibinfo{author}{Balents, L.}
\newblock \bibinfo{title}{{Correlated Quantum Phenomena in the Strong
  Spin-Orbit Regime}}.
\newblock \emph{\bibinfo{journal}{Annual Review of Condensed Matter Physics}}
  \textbf{\bibinfo{volume}{5}}, \bibinfo{pages}{57--82} (\bibinfo{year}{2014}).

\bibitem{Nakai2022}
\bibinfo{author}{Nakai, H.} \& \bibinfo{author}{Hotta, C.}
\newblock \bibinfo{title}{{Perfect flat band with chirality and charge ordering
  out of strong spin-orbit interaction}}.
\newblock \emph{\bibinfo{journal}{Nature Communications}}
  \textbf{\bibinfo{volume}{13}} (\bibinfo{year}{2022}).

\bibitem{Metzner1989}
\bibinfo{author}{Metzner, W.} \& \bibinfo{author}{Vollhardt, D.}
\newblock \bibinfo{title}{Correlated lattice fermions in
  $d=\ensuremath{\infty}$ dimensions}.
\newblock \emph{\bibinfo{journal}{Phys. Rev. Lett.}}
  \textbf{\bibinfo{volume}{62}}, \bibinfo{pages}{324--327}
  (\bibinfo{year}{1989}).

\bibitem{Georges1992}
\bibinfo{author}{Georges, A.} \& \bibinfo{author}{Kotliar, G.}
\newblock \bibinfo{title}{Hubbard model in infinite dimensions}.
\newblock \emph{\bibinfo{journal}{Phys. Rev. B}} \textbf{\bibinfo{volume}{45}},
  \bibinfo{pages}{6479--6483} (\bibinfo{year}{1992}).

\bibitem{Maier2005}
\bibinfo{author}{Maier, T.}, \bibinfo{author}{Jarrell, M.},
  \bibinfo{author}{Pruschke, T.} \& \bibinfo{author}{Hettler, M.~H.}
\newblock \bibinfo{title}{Quantum cluster theories}.
\newblock \emph{\bibinfo{journal}{Rev. Mod. Phys.}}
  \textbf{\bibinfo{volume}{77}}, \bibinfo{pages}{1027--1080}
  (\bibinfo{year}{2005}).

\bibitem{Kawano2022}
\bibinfo{author}{Kawano, M.} \& \bibinfo{author}{Hotta, C.}
\newblock \bibinfo{title}{{Sine-square deformed mean-field theory}}.
\newblock \emph{\bibinfo{journal}{Phys. Rev. Res.}}
  \textbf{\bibinfo{volume}{4}}, \bibinfo{pages}{L012033}
  (\bibinfo{year}{2022}).

\bibitem{Kawano2023}
\bibinfo{author}{Kawano, M.} \& \bibinfo{author}{Hotta, C.}
\newblock \bibinfo{title}{{Phase diagram of the square-lattice {Hubbard} model
  with {Rashba}-type antisymmetric spin-orbit coupling}}.
\newblock \emph{\bibinfo{journal}{Phys. Rev. B}}
  \textbf{\bibinfo{volume}{107}}, \bibinfo{pages}{045123}
  (\bibinfo{year}{2023}).

\bibitem{Knizia2012}
\bibinfo{author}{Knizia, G.} \& \bibinfo{author}{Chan, G. K.-L.}
\newblock \bibinfo{title}{{Density Matrix Embedding: A Simple Alternative to
  Dynamical Mean-Field Theory}}.
\newblock \emph{\bibinfo{journal}{Phys. Rev. Lett.}}
  \textbf{\bibinfo{volume}{109}}, \bibinfo{pages}{186404}
  (\bibinfo{year}{2012}).

\bibitem{Plat2020}
\bibinfo{author}{Plat, X.} \& \bibinfo{author}{Hotta, C.}
\newblock \bibinfo{title}{Entanglement spectrum as a marker for phase
  transitions in the density embedding theory for interacting spinless
  fermionic models}.
\newblock \emph{\bibinfo{journal}{Phys. Rev. B}}
  \textbf{\bibinfo{volume}{102}}, \bibinfo{pages}{140410}
  (\bibinfo{year}{2020}).

\bibitem{Kawano2020}
\bibinfo{author}{Kawano, M.} \& \bibinfo{author}{Hotta, C.}
\newblock \bibinfo{title}{Comparative study of the density matrix embedding
  theory for hubbard models}.
\newblock \emph{\bibinfo{journal}{Phys. Rev. B}}
  \textbf{\bibinfo{volume}{102}}, \bibinfo{pages}{235111}
  (\bibinfo{year}{2020}).

\bibitem{Hotta2013}
\bibinfo{author}{Hotta, C.}, \bibinfo{author}{Nishimoto, S.} \&
  \bibinfo{author}{Shibata, N.}
\newblock \bibinfo{title}{{Grand canonical finite size numerical approaches in
  one and two dimensions: {Real} space energy renormalization and edge state
  generation}}.
\newblock \emph{\bibinfo{journal}{Phys. Rev. B}} \textbf{\bibinfo{volume}{87}},
  \bibinfo{pages}{115128} (\bibinfo{year}{2013}).

\bibitem{Hotta2012}
\bibinfo{author}{Hotta, C.} \& \bibinfo{author}{Shibata, N.}
\newblock \bibinfo{title}{{Grand canonical finite-size numerical approaches: A
  route to measuring bulk properties in an applied field}}.
\newblock \emph{\bibinfo{journal}{Phys. Rev. B}} \textbf{\bibinfo{volume}{86}},
  \bibinfo{pages}{041108} (\bibinfo{year}{2012}).

\bibitem{Note1}
\bibinfo{note}{In standard mean-field approximation the magnetic insulators are
  over stabilized so that the UUD state shall appear at $B=0$ in both diagrams.
  However, SSDMF safely captures the paramagnetic ground state observed in
  previous studies taking account of the correlation effects at their maximum,
  indicating that it includes the higher-order correlation effects from the
  idea of energy renormalization done by SSD\cite {Hotta2013}. Indeed the
  120$^\circ $-based states appear in our phase diagram at the comparable
  $U/t\sim 10$ as Ref.[\protect \onlinecite {Shirakawa2017}].}

\bibitem{Morita2002}
\bibinfo{author}{Morita, H.}, \bibinfo{author}{Watanabe, S.} \&
  \bibinfo{author}{Imada, M.}
\newblock \bibinfo{title}{{Nonmagnetic insulating states near the Mott
  transitions on lattice with geometrical frustration and implications for
  $\kappa$-{ET}$_2${Cu}$_2$({CN})$_3$}}.
\newblock \emph{\bibinfo{journal}{J. Phys. Soc. Jpn.}}
  \textbf{\bibinfo{volume}{71}}, \bibinfo{pages}{2109} (\bibinfo{year}{2002}).

\bibitem{Yoshioka2009}
\bibinfo{author}{Yoshioka, T.}, \bibinfo{author}{Koga, A.} \&
  \bibinfo{author}{Kawakami, N.}
\newblock \bibinfo{title}{{Quantum Phase Transitions in the Hubbard Model on a
  Triangular Lattice}}.
\newblock \emph{\bibinfo{journal}{Phys. Rev. Lett.}}
  \textbf{\bibinfo{volume}{103}}, \bibinfo{pages}{036401}
  (\bibinfo{year}{2009}).

\bibitem{Shirakawa2017}
\bibinfo{author}{Shirakawa, T.}, \bibinfo{author}{Tohyama, T.},
  \bibinfo{author}{Kokalj, J.}, \bibinfo{author}{Sota, S.} \&
  \bibinfo{author}{Yunoki, S.}
\newblock \bibinfo{title}{{Ground-state phase diagram of the triangular lattice
  Hubbard model by the density-matrix renormalization group method}}.
\newblock \emph{\bibinfo{journal}{Phys. Rev. B}} \textbf{\bibinfo{volume}{96}},
  \bibinfo{pages}{205130} (\bibinfo{year}{2017}).

\bibitem{Sotnikov2021}
\bibinfo{author}{Sotnikov, O.~M.} \emph{et~al.}
\newblock \bibinfo{title}{{Probing the topology of the quantum analog of a
  classical skyrmion}}.
\newblock \emph{\bibinfo{journal}{Phys. Rev. B}}
  \textbf{\bibinfo{volume}{103}}, \bibinfo{pages}{L060404}
  (\bibinfo{year}{2021}).

\bibitem{Lohani2019}
\bibinfo{author}{Lohani, V.}, \bibinfo{author}{Hickey, C.},
  \bibinfo{author}{Masell, J.} \& \bibinfo{author}{Rosch, A.}
\newblock \bibinfo{title}{{Quantum Skyrmions in Frustrated Ferromagnets}}.
\newblock \emph{\bibinfo{journal}{Phys. Rev. X}} \textbf{\bibinfo{volume}{9}},
  \bibinfo{pages}{041063} (\bibinfo{year}{2019}).

\bibitem{Maeland2022}
\bibinfo{author}{M\ae{}land, K.} \& \bibinfo{author}{Sudb\o{}, A.}
\newblock \bibinfo{title}{{Quantum topological phase transitions in skyrmion
  crystals}}.
\newblock \emph{\bibinfo{journal}{Phys. Rev. Res.}}
  \textbf{\bibinfo{volume}{4}}, \bibinfo{pages}{L032025}
  (\bibinfo{year}{2022}).

\bibitem{Gao2017}
\bibinfo{author}{Gao, S.} \emph{et~al.}
\newblock \bibinfo{title}{{Spiral spin-liquid and the emergence of a
  vortex-like state in MnSc${}_{2}$S${}_{4}$}}.
\newblock \emph{\bibinfo{journal}{Nature Physics}}
  \textbf{\bibinfo{volume}{13}}, \bibinfo{pages}{157--161}
  (\bibinfo{year}{2017}).

\bibitem{Rosales2020}
\bibinfo{author}{Gao, S.} \emph{et~al.}
\newblock \bibinfo{title}{{Fractional antiferromagnetic skyrmion lattice
  induced by anisotropic couplings}}.
\newblock \emph{\bibinfo{journal}{Nature}} \textbf{\bibinfo{volume}{586}},
  \bibinfo{pages}{37--41} (\bibinfo{year}{2020}).

\bibitem{Rosales2022}
\bibinfo{author}{Mohylna, M.}, \bibinfo{author}{G\'omez~Albarrac\'{\i}n,
  F.~A.}, \bibinfo{author}{\ifmmode \check{Z}\else
  \v{Z}\fi{}ukovi\ifmmode~\check{c}\else \v{c}\fi{}, M.} \&
  \bibinfo{author}{Rosales, H.~D.}
\newblock \bibinfo{title}{{Spontaneous antiferromagnetic skyrmion/antiskyrmion
  lattice and spiral spin-liquid states in the frustrated triangular lattice}}.
\newblock \emph{\bibinfo{journal}{Phys. Rev. B}}
  \textbf{\bibinfo{volume}{106}}, \bibinfo{pages}{224406}
  (\bibinfo{year}{2022}).

\bibitem{Rosales2015}
\bibinfo{author}{Rosales, H.~D.}, \bibinfo{author}{Cabra, D.~C.} \&
  \bibinfo{author}{Pujol, P.}
\newblock \bibinfo{title}{{Three-sublattice skyrmion crystal in the
  antiferromagnetic triangular lattice}}.
\newblock \emph{\bibinfo{journal}{Phys. Rev. B}} \textbf{\bibinfo{volume}{92}},
  \bibinfo{pages}{214439} (\bibinfo{year}{2015}).

\bibitem{Takeda2024}
\bibinfo{author}{Takeda, H.} \emph{et~al.}
\newblock \bibinfo{title}{{Magnon thermal Hall effect via emergent SU(3) flux
  on the antiferromagnetic skyrmion lattice}}.
\newblock \emph{\bibinfo{journal}{Nature Communications}}
  \textbf{\bibinfo{volume}{15}}, \bibinfo{pages}{566} (\bibinfo{year}{2024}).

\bibitem{LeBlanc2015}
\bibinfo{author}{LeBlanc, J. P.~F.} \emph{et~al.}
\newblock \bibinfo{title}{Solutions of the two-dimensional hubbard model:
  Benchmarks and results from a wide range of numerical algorithms}.
\newblock \emph{\bibinfo{journal}{Phys. Rev. X}} \textbf{\bibinfo{volume}{5}},
  \bibinfo{pages}{041041} (\bibinfo{year}{2015}).

\bibitem{Fukui2005}
\bibinfo{author}{Fukui, T.}, \bibinfo{author}{Hatsugai, Y.} \&
  \bibinfo{author}{Suzuki, H.}
\newblock \bibinfo{title}{{Chern Numbers in Discretized Brillouin Zone:
  {Efficient} Method of Computing (Spin) Hall Conductances}}.
\newblock \emph{\bibinfo{journal}{J. Phys. Soc. Jpn.}}
  \textbf{\bibinfo{volume}{74}}, \bibinfo{pages}{1674--1677}
  (\bibinfo{year}{2005}).

\bibitem{Kaplan1983}
\bibinfo{author}{Kaplan, T.~A.}
\newblock \bibinfo{title}{{Single-band Hubbard model with spin-orbit
  coupling}}.
\newblock \emph{\bibinfo{journal}{Zeitschrift f{\"u}r Physik B Condensed
  Matter}} \textbf{\bibinfo{volume}{49}}, \bibinfo{pages}{313--317}
  (\bibinfo{year}{1983}).

\bibitem{Shekhtman1992}
\bibinfo{author}{Shekhtman, L.}, \bibinfo{author}{Entin-Wohlman, O.} \&
  \bibinfo{author}{Aharony, A.}
\newblock \bibinfo{title}{Moriya's anisotropic superexchange interaction,
  frustration, and dzyaloshinsky's weak ferromagnetism}.
\newblock \emph{\bibinfo{journal}{Phys. Rev. Lett.}}
  \textbf{\bibinfo{volume}{69}}, \bibinfo{pages}{836--839}
  (\bibinfo{year}{1992}).

\bibitem{Shekhtman1993}
\bibinfo{author}{Shekhtman, L.}, \bibinfo{author}{Aharony, A.} \&
  \bibinfo{author}{Entin-Wohlman, O.}
\newblock \bibinfo{title}{{Bond-dependent symmetric and antisymmetric
  superexchange interactions in La${}_{2}$CuO${}_{4}$}}.
\newblock \emph{\bibinfo{journal}{Phys. Rev. B}} \textbf{\bibinfo{volume}{47}},
  \bibinfo{pages}{174--182} (\bibinfo{year}{1993}).

\bibitem{Nayak2017}
\bibinfo{author}{Nayak, A.~K.} \emph{et~al.}
\newblock \bibinfo{title}{{Magnetic antiskyrmions above room temperature in
  tetragonal Heusler materials}}.
\newblock \emph{\bibinfo{journal}{Nature}} \textbf{\bibinfo{volume}{548}},
  \bibinfo{pages}{561--566} (\bibinfo{year}{2017}).

\bibitem{Peng2020}
\bibinfo{author}{Peng, L.} \emph{et~al.}
\newblock \bibinfo{title}{{Controlled transformation of skyrmions and
  antiskyrmions in a non-centrosymmetric magnet}}.
\newblock \emph{\bibinfo{journal}{Nature Nanotechnology}}
  \textbf{\bibinfo{volume}{15}}, \bibinfo{pages}{181--186}
  (\bibinfo{year}{2020}).

\bibitem{Shibata2013}
\bibinfo{author}{Shibata, K.} \emph{et~al.}
\newblock \bibinfo{title}{{Towards control of the size and helicity of
  skyrmions in helimagnetic alloys by spin--orbit coupling}}.
\newblock \emph{\bibinfo{journal}{Nature Nanotechnology}}
  \textbf{\bibinfo{volume}{8}}, \bibinfo{pages}{723--728}
  (\bibinfo{year}{2013}).

\bibitem{Grigoriev2013}
\bibinfo{author}{Grigoriev, S.~V.} \emph{et~al.}
\newblock \bibinfo{title}{{Chiral Properties of Structure and Magnetism in
  ${\mathrm{Mn}}_{1\mathrm{\text{\ensuremath{-}}}x}{\mathrm{Fe}}_{x}\mathrm{Ge}$
  Compounds: When the Left and the Right are Fighting, Who Wins?}}
\newblock \emph{\bibinfo{journal}{Phys. Rev. Lett.}}
  \textbf{\bibinfo{volume}{110}}, \bibinfo{pages}{207201}
  (\bibinfo{year}{2013}).

\bibitem{Kanazawa2011}
\bibinfo{author}{Kanazawa, N.} \emph{et~al.}
\newblock \bibinfo{title}{{Large Topological Hall Effect in a Short-Period
  Helimagnet {MnGe}}}.
\newblock \emph{\bibinfo{journal}{Phys. Rev. Lett.}}
  \textbf{\bibinfo{volume}{106}}, \bibinfo{pages}{156603}
  (\bibinfo{year}{2011}).

\bibitem{Tanigaki2015}
\bibinfo{author}{Tanigaki, T.} \emph{et~al.}
\newblock \bibinfo{title}{{Real-Space Observation of Short-Period Cubic Lattice
  of Skyrmions in {MnGe}}}.
\newblock \emph{\bibinfo{journal}{Nano Letters}} \textbf{\bibinfo{volume}{15}},
  \bibinfo{pages}{5438--5442} (\bibinfo{year}{2015}).

\bibitem{Adams2012}
\bibinfo{author}{Adams, T.} \emph{et~al.}
\newblock \bibinfo{title}{{Long-Wavelength Helimagnetic Order and Skyrmion
  Lattice Phase in ${\mathrm{Cu}}_{2}{\mathrm{OSeO}}_{3}$}}.
\newblock \emph{\bibinfo{journal}{Phys. Rev. Lett.}}
  \textbf{\bibinfo{volume}{108}}, \bibinfo{pages}{237204}
  (\bibinfo{year}{2012}).

\bibitem{Fujishiro2019}
\bibinfo{author}{Fujishiro, Y.} \emph{et~al.}
\newblock \bibinfo{title}{{Topological transitions among skyrmion- and
  hedgehog-lattice states in cubic chiral magnets}}.
\newblock \emph{\bibinfo{journal}{Nature Communications}}
  \textbf{\bibinfo{volume}{10}}, \bibinfo{pages}{1059} (\bibinfo{year}{2019}).

\bibitem{Karube2018}
\bibinfo{author}{Karube, K.} \emph{et~al.}
\newblock \bibinfo{title}{{Controlling the helicity of magnetic skyrmions in a
  $\ensuremath{\beta}$-{Mn}-type high-temperature chiral magnet}}.
\newblock \emph{\bibinfo{journal}{Phys. Rev. B}} \textbf{\bibinfo{volume}{98}},
  \bibinfo{pages}{155120} (\bibinfo{year}{2018}).

\bibitem{Li2016}
\bibinfo{author}{Li, W.} \emph{et~al.}
\newblock \bibinfo{title}{{Emergence of skyrmions from rich parent phases in
  the molybdenum nitrides}}.
\newblock \emph{\bibinfo{journal}{Phys. Rev. B}} \textbf{\bibinfo{volume}{93}},
  \bibinfo{pages}{060409} (\bibinfo{year}{2016}).

\bibitem{Kaneko2019}
\bibinfo{author}{Kaneko, K.} \emph{et~al.}
\newblock \bibinfo{title}{{Unique Helical Magnetic Order and Field-Induced
  Phase in Trillium Lattice Antiferromagnet EuPtSi}}.
\newblock \emph{\bibinfo{journal}{Journal of the Physical Society of Japan}}
  \textbf{\bibinfo{volume}{88}}, \bibinfo{pages}{013702}
  (\bibinfo{year}{2019}).

\bibitem{Kakihana2018}
\bibinfo{author}{Kakihana, M.} \emph{et~al.}
\newblock \bibinfo{title}{{Giant Hall Resistivity and Magnetoresistance in
  Cubic Chiral Antiferromagnet EuPtSi}}.
\newblock \emph{\bibinfo{journal}{Journal of the Physical Society of Japan}}
  \textbf{\bibinfo{volume}{87}}, \bibinfo{pages}{023701}
  (\bibinfo{year}{2018}).

\bibitem{Heinze2011}
\bibinfo{author}{Heinze, S.} \emph{et~al.}
\newblock \bibinfo{title}{{Spontaneous atomic-scale magnetic skyrmion lattice
  in two dimensions}}.
\newblock \emph{\bibinfo{journal}{Nature Physics}}
  \textbf{\bibinfo{volume}{7}}, \bibinfo{pages}{713--718}
  (\bibinfo{year}{2011}).

\bibitem{Romming2013}
\bibinfo{author}{Romming, N.} \emph{et~al.}
\newblock \bibinfo{title}{{Writing and Deleting Single Magnetic Skyrmions}}.
\newblock \emph{\bibinfo{journal}{Science}} \textbf{\bibinfo{volume}{341}},
  \bibinfo{pages}{636--639} (\bibinfo{year}{2013}).

\bibitem{Moreau-Luchaire2016}
\bibinfo{author}{Moreau-Luchaire, C.} \emph{et~al.}
\newblock \bibinfo{title}{{Additive interfacial chiral interaction in
  multilayers for stabilization of small individual skyrmions at room
  temperature}}.
\newblock \emph{\bibinfo{journal}{Nature Nanotechnology}}
  \textbf{\bibinfo{volume}{11}}, \bibinfo{pages}{444--448}
  (\bibinfo{year}{2016}).

\bibitem{Yasuda2016}
\bibinfo{author}{Yasuda, K.} \emph{et~al.}
\newblock \bibinfo{title}{{Geometric Hall effects in topological
  insulator heterostructures}}.
\newblock \emph{\bibinfo{journal}{Nature Physics}}
  \textbf{\bibinfo{volume}{12}}, \bibinfo{pages}{555--559}
  (\bibinfo{year}{2016}).

\bibitem{Matsuno2016}
\bibinfo{author}{Matsuno, J.} \emph{et~al.}
\newblock \bibinfo{title}{{Interface-driven topological {Hall} effect in
  {SrRuO}$_3$-{SrIrO}$_3$ bilayer}}.
\newblock \emph{\bibinfo{journal}{Science Advances}}
  \textbf{\bibinfo{volume}{2}} (\bibinfo{year}{2016}).

\bibitem{Wang2018}
\bibinfo{author}{Wang, L.} \emph{et~al.}
\newblock \bibinfo{title}{{Ferroelectrically tunable magnetic skyrmions in
  ultrathin oxide heterostructures}}.
\newblock \emph{\bibinfo{journal}{Nature Materials}}
  \textbf{\bibinfo{volume}{17}}, \bibinfo{pages}{1087--1094}
  (\bibinfo{year}{2018}).

\bibitem{Bordacs2017}
\bibinfo{author}{Bord{\'a}cs, S.} \emph{et~al.}
\newblock \bibinfo{title}{{Equilibrium Skyrmion Lattice Ground State in a Polar
  Easy-plane Magnet}}.
\newblock \emph{\bibinfo{journal}{Scientific Reports}}
  \textbf{\bibinfo{volume}{7}}, \bibinfo{pages}{7584} (\bibinfo{year}{2017}).

\bibitem{Ozawa2017}
\bibinfo{author}{Ozawa, R.}, \bibinfo{author}{Hayami, S.} \&
  \bibinfo{author}{Motome, Y.}
\newblock \bibinfo{title}{{Zero-Field Skyrmions with a High Topological Number
  in Itinerant Magnets}}.
\newblock \emph{\bibinfo{journal}{Phys. Rev. Lett.}}
  \textbf{\bibinfo{volume}{118}}, \bibinfo{pages}{147205}
  (\bibinfo{year}{2017}).

\bibitem{Choi2019}
\bibinfo{author}{Choi, H.}, \bibinfo{author}{Tai, Y.~Y.} \&
  \bibinfo{author}{Zhu, J.~X.}
\newblock \bibinfo{title}{{Spin-fermion model for skyrmions in MnGe derived
  from strong correlations}}.
\newblock \emph{\bibinfo{journal}{Phys. Rev. B}} \textbf{\bibinfo{volume}{99}},
  \bibinfo{pages}{134437} (\bibinfo{year}{2019}).

\bibitem{Kathyat2020}
\bibinfo{author}{Kathyat, D.~S.}, \bibinfo{author}{Mukherjee, A.} \&
  \bibinfo{author}{Kumar, S.}
\newblock \bibinfo{title}{{Microscopic magnetic Hamiltonian for exotic spin
  textures in metals}}.
\newblock \emph{\bibinfo{journal}{Phys. Rev. B}}
  \textbf{\bibinfo{volume}{102}}, \bibinfo{pages}{075106}
  (\bibinfo{year}{2020}).

\bibitem{Mukherjee2021}
\bibinfo{author}{Mukherjee, A.}, \bibinfo{author}{Kathyat, D.~S.} \&
  \bibinfo{author}{Kumar, S.}
\newblock \bibinfo{title}{{Antiferromagnetic skyrmions and skyrmion density
  wave in a Rashba-coupled Hund insulator}}.
\newblock \emph{\bibinfo{journal}{Phys. Rev. B}}
  \textbf{\bibinfo{volume}{103}}, \bibinfo{pages}{134424}
  (\bibinfo{year}{2021}).

\bibitem{Kathyat2021}
\bibinfo{author}{Kathyat, D.~S.}, \bibinfo{author}{Mukherjee, A.} \&
  \bibinfo{author}{Kumar, S.}
\newblock \bibinfo{title}{{Electronic mechanism for nanoscale skyrmions and
  topological metals}}.
\newblock \emph{\bibinfo{journal}{Phys. Rev. B}}
  \textbf{\bibinfo{volume}{103}}, \bibinfo{pages}{035111}
  (\bibinfo{year}{2021}).

\bibitem{Mukherjee2022}
\bibinfo{author}{Mukherjee, A.}, \bibinfo{author}{Kathyat, D.~S.} \&
  \bibinfo{author}{Kumar, S.}
\newblock \bibinfo{title}{Engineering antiferromagnetic skyrmions and
  antiskyrmions at metallic interfaces}.
\newblock \emph{\bibinfo{journal}{Phys. Rev. B}}
  \textbf{\bibinfo{volume}{105}}, \bibinfo{pages}{075102}
  (\bibinfo{year}{2022}).

\bibitem{Mukherjee2023}
\bibinfo{author}{Mukherjee, A.}, \bibinfo{author}{Sanyal, A.~B.} \&
  \bibinfo{author}{Dagotto, E.}
\newblock \bibinfo{title}{Unconventional skyrmions in an interfacial frustrated
  triangular lattice}.
\newblock \emph{\bibinfo{journal}{Phys. Rev. B}}
  \textbf{\bibinfo{volume}{108}}, \bibinfo{pages}{014408}
  (\bibinfo{year}{2023}).

\bibitem{Koelling1977}
\bibinfo{author}{Koelling, D.~D.} \& \bibinfo{author}{Harmon, B.}
\newblock \bibinfo{title}{{A technique for relativistic spin-polarised
  calculations}}.
\newblock \emph{\bibinfo{journal}{Journal of Physics C: Solid State Physics}}
  \textbf{\bibinfo{volume}{10}}, \bibinfo{pages}{3107--3114}
  (\bibinfo{year}{1977}).

\bibitem{kurita2011}
\bibinfo{author}{Kurita, M.}, \bibinfo{author}{Yamaji, Y.} \&
  \bibinfo{author}{Imada, M.}
\newblock \bibinfo{title}{{Topological Insulators from Spontaneous Symmetry
  Breaking Induced by Electron Correlation on Pyrochlore Lattices}}.
\newblock \emph{\bibinfo{journal}{J. Phys. Soc. Jpn.}}
  \textbf{\bibinfo{volume}{80}}, \bibinfo{pages}{044708}
  (\bibinfo{year}{2011}).

\bibitem{Shibata2011}
\bibinfo{author}{Shibata, N.} \& \bibinfo{author}{Hotta, C.}
\newblock \bibinfo{title}{Boundary effects in the density-matrix
  renormalization group calculation}.
\newblock \emph{\bibinfo{journal}{Phys. Rev. B}} \textbf{\bibinfo{volume}{84}},
  \bibinfo{pages}{115116} (\bibinfo{year}{2011}).

\bibitem{Hotta2012crystal}
\bibinfo{author}{Hotta, C.}
\newblock \bibinfo{title}{Theories on frustrated electrons in two-dimensional
  organic solids}.
\newblock \emph{\bibinfo{journal}{Crystals}} \textbf{\bibinfo{volume}{2}},
  \bibinfo{pages}{1155--1200} (\bibinfo{year}{2012}).

\bibitem{Haller2022}
\bibinfo{author}{Haller, A.}, \bibinfo{author}{Groenendijk, S.},
  \bibinfo{author}{Habibi, A.}, \bibinfo{author}{Michels, A.} \&
  \bibinfo{author}{Schmidt, T.~L.}
\newblock \bibinfo{title}{{Quantum skyrmion lattices in Heisenberg
  ferromagnets}}.
\newblock \emph{\bibinfo{journal}{Phys. Rev. Res.}}
  \textbf{\bibinfo{volume}{4}}, \bibinfo{pages}{043113} (\bibinfo{year}{2022}).

\bibitem{Takahashi1977}
\bibinfo{author}{Takahashi, M.}
\newblock \bibinfo{title}{{Half-filled {Hubbard} model at low temperature}}.
\newblock \emph{\bibinfo{journal}{J. Phys. C}} \textbf{\bibinfo{volume}{10}},
  \bibinfo{pages}{1289--7301} (\bibinfo{year}{1977}).

\end{thebibliography}
\end{document}